\documentclass[%
pre,preprint,
superscriptaddress,
tightenlines,
showpacs,showkeys,
twocolumn,
a4paper,12pt
]{revtex4}
\usepackage{dcolumn}
\usepackage{amssymb,amsmath,array}
\usepackage{graphicx}
\usepackage{subfigure}

\renewcommand{\thesubfigure}{(\alph{subfigure})}

\makeatletter
\renewcommand{\@thesubfigure}{\thesubfigure\space}
 
\newcommand{\bs}[1]{\boldsymbol{#1}}
\newcommand{\vc}[1]{\mathbf{#1}}
\newcommand{\uvc}[1]{\mathbf{\hat #1}}

\newcommand{\dd}{\mathrm{d}}
\newcommand{\pdr}[1]{\frac{\partial #1}{\partial t}}

\newcommand{\trans}{\emph{trans}}
\newcommand{\cis}{\emph{cis}}
\newcommand{\degc}{$^o$C}

\makeatother

\begin{document}
\DeclareGraphicsExtensions{.jpg,.pdf}
\title{Photoinduced 3D orientational order in side chain liquid
  crystalline azopolymers}

 \author{O.V.~Yaroshchuk}
 \email[Email address: ]{olegyar@iop.kiev.ua}
 \affiliation{%
 Institute of Physics of NASU, pr. Nauki 46,
 03028 Ky\"{\i}v, Ukraine
 }

\author{A.D.~Kiselev}
\email[Email address: ]{kisel@elit.chernigov.ua}
\affiliation{%
 Chernigov State Technological University,
 Shevchenko Street 95,
 14027 Chernigov, Ukraine
} 

\author{Yu. Zakrevskyy}
\affiliation{%
 Institute of Physics of NASU, pr. Nauki 46,
 03028 Ky\"{\i}v, Ukraine
 } 

\author{T. Bidna}
\affiliation{%
 Institute of Physics of NASU, pr. Nauki 46,
 03028 Ky\"{\i}v, Ukraine
 }

\author{J. Kelly}
\affiliation{%
Kent State University, Liquid Crystal Institute,
44242-0001 Kent, Ohio, USA
}

\author{L.-C. Chien}
\affiliation{%
Kent State University, Liquid Crystal Institute,
44242-0001 Kent, Ohio, USA
}

\author{J. Lindau}
\affiliation{%
Institute of Physical Chemistry,
Martin-Luther University, M\"{u}hlphorte 1, 06106 Halle, Germany
}

\date{\today}

\begin{abstract}
  We apply experimental technique based on the combination of methods
  dealing with principal refractive indices and absorption
  coefficients to study the photoinduced 3D orientational order in the
  films of liquid crystalline (LC) azopolymers.  The technique is used
  to identify 3D orientational configurations of \trans\ azobenzene
  chromophores and to characterize the degree of ordering in terms of
  order parameters.  We study two types of LC azopolymers which form
  structures with preferred in-plane and out-of-plane alignment of
  azochromophores, correspondingly.  Using irradiation with the
  polarized light of two different wavelengths we find that the
  kinetics of photoinduced anisotropy can be dominated by either
  photo-reorientation (angular redistribution of \trans\ chromophores)
  or photoselection (angular selective \trans-\cis\ isomerization)
  mechanisms depending on the wavelength.  At early stages of
  irradiation the films of both azopolymers are biaxial. This
  biaxiality disappears on reaching a state of photo-saturation. In
  the regime of photoselection the photo-saturated state of the film
  is optically isotropic.  But, in the case of the photo-reorientation
  mechanism, anisotropy of this state is uniaxial with the optical
  axis dependent on the preferential alignment of azochromophores.  We
  formulate the phenomenological model describing the kinetics of
  photoinduced anisotropy in terms of the isomer concentrations and
  the order parameter tensor. We present the numerical results for
  absorption coefficients that are found to be in good agreement with
  the experimental data.  The model is also used to interpret the
  effect of changing the mechanism with the wavelength of the pumping
  light.

\end{abstract}

 
\maketitle

\section{Introduction}
\label{sec:intro}

Azobenzene and its derivatives have been attracted much attention over
the past few decades because of a number of fascinating features of
these compounds. They were initially used in preparation of paints,
because of rich spectrum of colors that can be obtained depending on
the chemical structure of azochromophores. Further investigations
revealed strong photochromism (photomodification of color) of
azobenzene derivatives.  It was found that this effect in azobenzene
derivatives can be accompanied by a novel phenomenon that is known as
photoinduced optical anisotropy (POA) and can be detected in dichroism
and birefringence measurements~\cite{Wieg:1919}.

The discovery of POA  opened up a new chapter 
in studies of azobenzene compounds. 
Neporent and
Stolbova~\cite{Nep:1963} described POA  in viscous
solutions of azodyes, then Todorov and co-workers~\cite{Tod:1984}
disclosed the same phenomenon in azodye-polymer blends. The anisotropy
induced in these systems is rather unstable. The stable POA was
observed later on in polymers containing chemically linked
azochromophores (azopolymers)~\cite{Eich:1987}.  It turned out that
stable anisotropy can be induced in both amorphous and liquid
crystalline (LC)
azopolymers~\cite{Eich:1987,Nat:1992,Holme:1996,Petry:1993,Wies:1992,Blin:1998,Yar:2001}.
The efficiency of POA in LC azopolymers is generally much higher than
it is for amorphous homologues. For example, the photoinduced
birefringence can be as high as 0.3~\cite{Yar:2001} that is a typical
value for low-molecular-weight LCs~\cite{Blin:b:1994}.

It is
commonly accepted that the macroscopic anisotropy detected in optical
experiments reflects the microscopic orientational order of
polymer fragments which is mainly determined by the order of azochromophores in
\trans\  configuration. There are two known phenomena 
underlying the process of orientational ordering of the azochromophores: 
(a)~strong absorption dichroism of azobenzene groups 
that have the optical transition dipole moment 
approximately directed along the long molecular axis; 
(b)~photochemically induced \trans-\cis\  isomerization and subsequent
thermal and/or photochemical \cis-\trans\  back isomerization of
azobenzene moieties. 

It means that the rate of the photoinduced isomerization strongly
depends on orientation of the azobenzene chromophores relative
to the polarization vector of actinic light $\vc{E}$.  The
fragments oriented perpendicular to $\vc{E}$ are almost inactive,
whereas the groups with the long axes parallel to $\vc{E}$ are
the most active for isomerization.

There are two different regimes of the photoinduced ordering.
These regimes are usually recognized 
as two limiting cases related to the lifetime of
\cis\  isomers under irradiation, and were theoretically considered
in Refs.~\cite{Dum:1992,Dum:spie:1994}. 
 
If \cis\ isomers are long-living,
the anisotropy is caused by angular selective
burning of mesogenic \trans\  isomers 
due to stimulated transitions to
non-mesogenic \cis\  form. 
The transition rate is angular dependent, so that
the \trans\  azochromophores normally oriented to $\vc{E}$ 
are the least burned. This
direction will define the axis of the induced anisotropy. 
This regime of POA 
is known as the mechanism of angular hole burning (photoselection).

In the opposite case of short-living \cis\ isomers, the azochromophores are
excited many times during the POA generation period.  These
\trans-\cis-\trans\ isomerization cycles are accompanied by 
rotations of the azochromophores that tend to 
assume orientation transverse
to $\vc{E}$ and minimize the
absorption. Non-photosensitive groups can be involved in the process
of reorientation due to cooperative
motion~\cite{Holme:1996,Puch:1998}.  This regime is referred to as the
photo-reorientation (angular redistribution) mechanism. 
  
From the above it might be concluded that, whichever mechanism of the
ordering is assumed, the exciting light causes preferential alignment
of azobenzene chromophores along the directions perpendicular to the
polarization vector of the actinic light. In 3D space these directions
form the plane normal to $\vc{E}$.  It can be expected that the
angular distribution of azochromophores in this plane is isotropic.
From the experimental results, however, this is not the case.  The
uniaxial ordering with strongly preferred in-plane alignment was
observed in Refs.~\cite{Yar:2001,Zakr:2001}.  In addition, it was
found that the photoinduced orientational structures can show
biaxiality~\cite{Wies:1992,Buff:1998,Kis:cond:2001,Yar:2001,Kis:epj:2001}.
 
This implies that actually the system is not spherically symmetric
in the absence of light and 
some of the above directions are preferred 
depending on a number of additional factors such as
irradiation conditions, chemical structure of
polymers, surface interaction etc. 
These factors combined with the action of light
may result in a large variety of orientational
configurations (uniaxial, biaxial, splayed) characterized by different
spatial orientations of the principal axes.  

In the past years this spatial character of the photoinduced anisotropies
has not received much attention and,  until recently,
it has been neglected in the bulk of experimental and theoretical
studies of POA in azobenzene containing
polymers~\cite{Eich:1987,Hvil:1992,Nat:1998,Dum:1992,Dum:spie:1994,Stu:1994,
  Ped:1997,Ped:1998,Puch:1998,Puch:1999}. 
On the other hand, 
the problems related to the 3D orientational structures in
polymeric films are currently of considerable importance in
the development of new compensation films for liquid crystal (LC)
displays~\cite{Serg:1999} and the pretilt angle generation by the use
of photoalignment method of LC orientation~\cite{Dyad:1995}.
  
The known
methods suitable for the experimental study of the 3D orientational
distributions in polymer films can be divided into two groups.

The methods of the first group are based on absorption
measurements. These methods have the indisputable advantage that the
order parameters of various molecular groups can be estimated from the
results of these measurements.  Shortcomings of the known absorption
methods~\cite{Wies:1992,Blin:1983} are the limited field of
applications and the strong approximations.
 
The second group includes the methods dealing with principal
refractive indices.  Recently several variations of the prism coupling
methods have been applied to measure the principal refractive indices
in azopolymer films~\cite{Osm:1999,Feng:1995,Cim:1999}.  These
results, however, were not used for in-depth analysis of such features
of the spatial ordering as biaxiality and spatial orientation of the
optical axes depending on polymer chemical structure, irradiation
conditions etc.

Our goal is a comprehensive investigation of the peculiarities of 3D 
orientational ordering in azopolymers.

In our previous studies~\cite{Kis:epj:2001,Yar:2001,Zakr:2001} we were
mainly concerned with the effects related to the peculiarities of
azopolymer self-organization that, in addition to the symmetry axis
defined by the light polarization, affect the light-induced 3D
ordering.  The combination of methods described in
Ref.~\cite{Kis:epj:2001} was found to be an experimental technique
particularly suitable to characterize light induced anisotropy
of orientational
structures in azopolymer films.  We have investigated these structures
depending on the polarization state of the actinic
light~\cite{Yar:2001} and on the molecular
constitution~\cite{Zakr:2001}.  
It was additionally found that the kinetics of POA
involving both uniaxial and biaxial structures can be theoretically
described by using the phenomenological approach suggested in
Ref.~\cite{Kis:epj:2001}.

In this work we concentrate on the features of the 3D orientation
determined by the discussed above mechanisms of the photo-induced
ordering. The different regimes of POA are realized experimentally by
choosing appropriate polymers and irradiation conditions. There are
two azopolymers of different structure used in this study and the
optical anisotropy was induced by irradiating the samples at two
different wavelengths. We find that the regime of POA in one of the
polymers strongly depends on the wavelength of the exciting light,
whereas the other presents the case in which POA is governed by the
photo-reorientation mechanism regardless of the wavelength.

We show that the experimental results can be interpreted on the basis
of the phenomenological model describing the kinetics of POA in terms
of angular redistribution probabilities and order parameter
correlation functions. This model can be deduced by using 
the procedure of Refs.~\cite{Kis:epj:2001,Kis:jpcm:2002} 
and is found to give the results that are in
good agreement with the experimental data. We also apply the model to
calculate the out-of-plane absorbance that cannot be directly estimated
from the results of measurements in the regime of photoselection.

The paper is organized as follows.  Sec.~\ref{sec:experiment} contains
the details on the combination of null ellipsometry and absorption
methods used in this study as an experimental procedure.  The
experimentally measured dependencies of birefringence and absorption
dichroism on the illumination doses are presented in
Sec.~\ref{sec:exper-results}.  We draw together these results to
unambiguously identify the anisotropy of the orientational 
structures induced in
azopolymer films and to measure the ordering of azochromophores
through the absorption order parameter. Anisotropy of the initial
structures and of the structures in the regime of photosaturation is
found to be uniaxial in both azopolymers. At early stages of
irradiation the transient anisotropic structures are biaxial.  It is
shown that under certain conditions the regime of the kinetics of POA
can be changed with the wavelength of the pumping light.
We also discuss
thermal stability of the photoinduced anisotropy
and experimental estimates of the photochemical parameters
needed for theoretical calculations.
In Sec.~\ref{sec:model} we describe the theoretical model 
formulated by using the phenomenological approach
of~\cite{Kis:epj:2001,Kis:jpcm:2002}. 
We discuss the physical assumptions underlying the model and
make brief comments on the
derivation of the kinetic equations for the concentrations of isomers
and the order parameter tensor.
In Sec.~\ref{sec:num-res} we present numerical results for
the absorption coefficients and the order parameters in relation to
the irradiation dose calculated by solving the equations of the model.
We compare these results with the experimental data
and comment on the predictions of the model concerning biaxiality effects,
stability and the regimes of POA.
Finally, general discussion of our results and 
some concluding remarks are given in Sec.~\ref{sec:discussion}.  

\begin{figure*}[!thb]
\centering
\subfigure[Polymer P1]{%
\resizebox{93mm}{!}{\includegraphics*{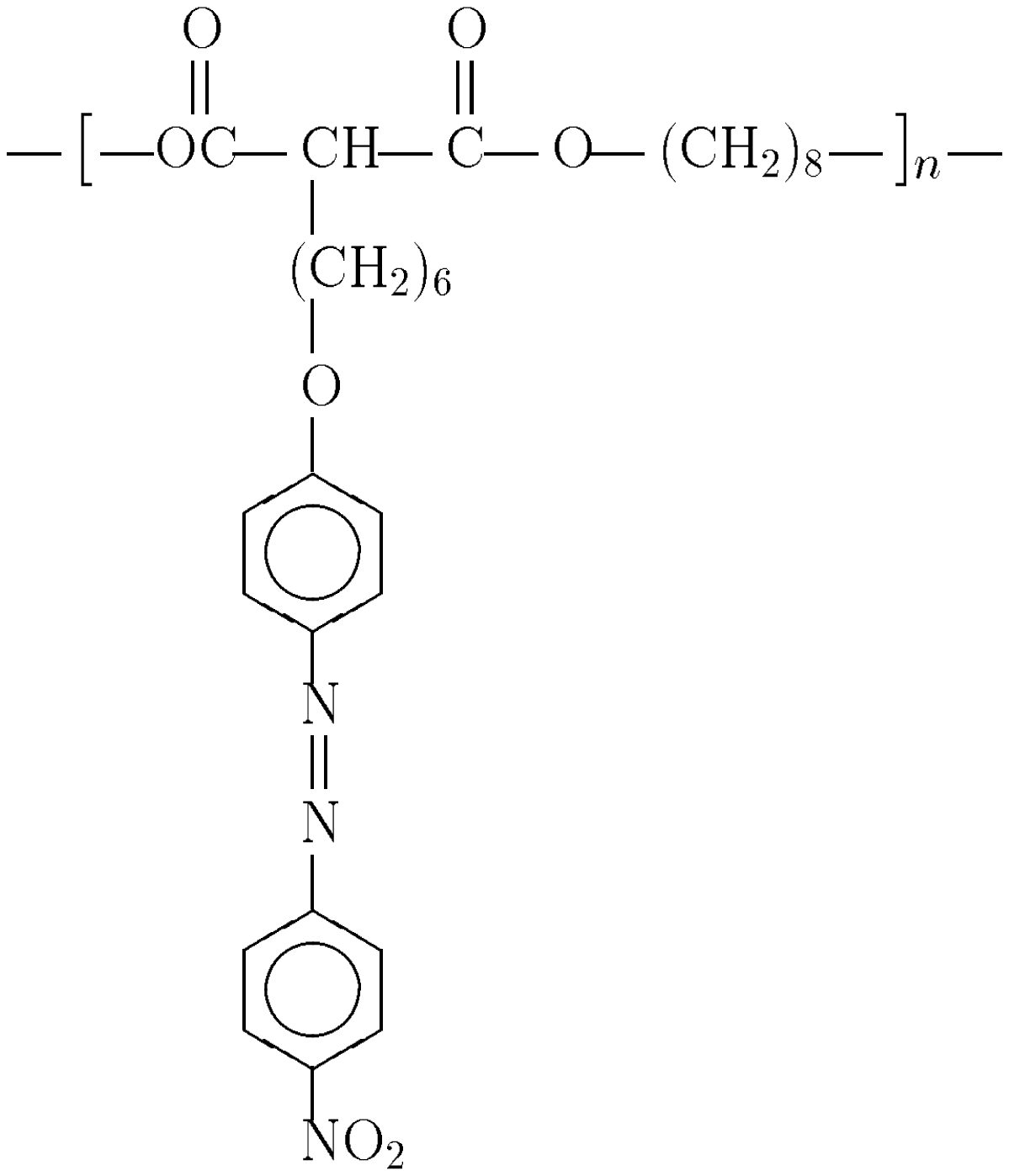}}
\label{fig:chem_p1}}
\subfigure[Polymer P2]{%
\resizebox{35mm}{!}{\includegraphics*{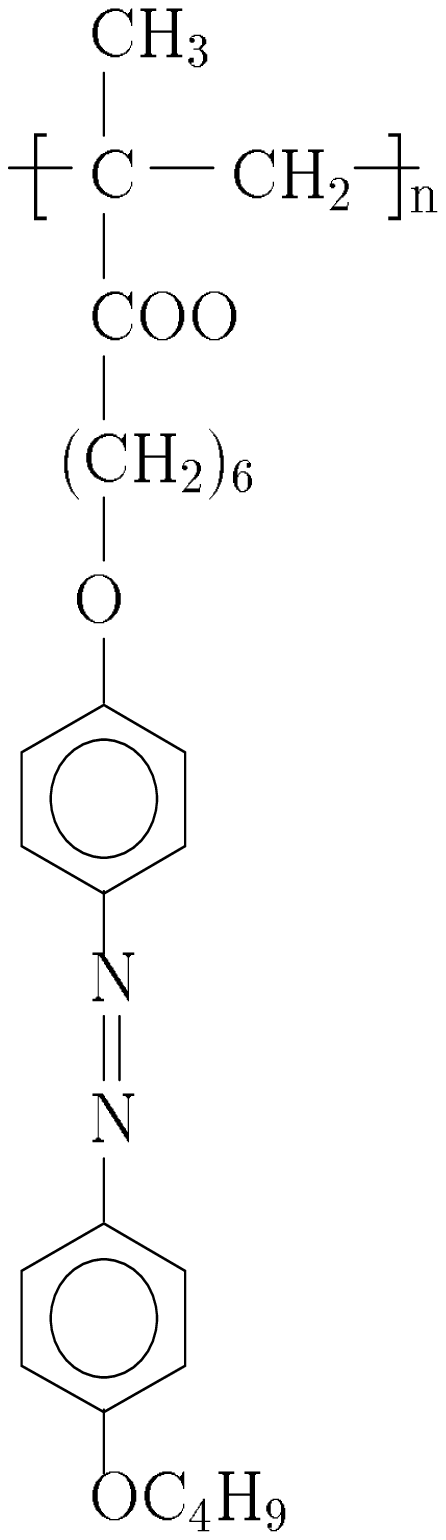}}
\label{fig:chem_p2}}
\caption{
Chemical structure of polymers.
}
\label{fig:chem}
\end{figure*}

\section{Experimental procedure}
\label{sec:experiment}

\subsection{Polymers}
\label{subsec:polymers}

The chemical formulas of two azopolymers used in our experiments are
presented in Fig.~\ref{fig:chem}.  
The details on synthesis of the polymers P1 and
P2 can be found in Ref.~\cite{Bohm:1993} and in
Ref.~\cite{Yar:c:2001}, correspondingly.  The polymers were
characterized by elemental analysis and 1H NMR spectroscopy. Molecular
weight of the polymers was determined by gel permeation
chromatography. The average molecular weight of P1 and P2 is estimated
at 13,500 g/mol and 36,000 g/mol, respectively.  Both materials are
comb-like polymers with azobenzene fragments in side chains connected
by flexible alkyl spacers to the polymer backbone. The azobenzene
chromophore of the polymer P1 contains polar $NO_2$ end group that has
strong acceptor properties. The azobenzene chromophore of polymer P2
contains hydrophobic alkyl tail $OC_{4}H_{9}$. The phase transitions
in the polymers were investigated by differential scanning calorimetry
and polarization microscopy.  Both azopolymers are found to possess
liquid crystalline properties in the respective temperature regions.
Polymer P1 forms smectic A and nematic mesophases within the
temperature range 44\degc--52\degc\ and 52\degc--55\degc,
respectively. Polymer P2 has a nematic mesophase within the
temperature interval 112\degc--140\degc. Both polymers are solids at
room temperature.

\begin{figure*}[!tbh]
\centering
\resizebox{150mm}{!}{\includegraphics*{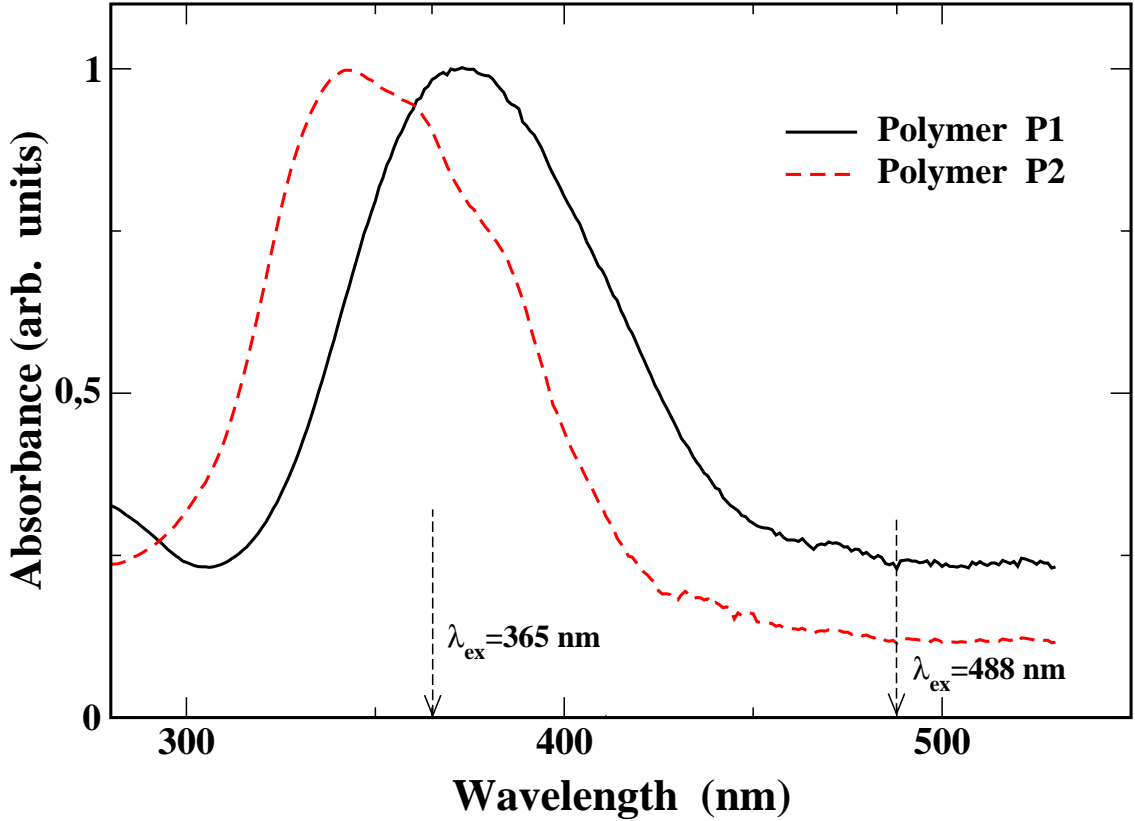}}
\caption{%
UV/Vis  absorption spectra of polymer films.  
}
\label{fig:UV-abs}
\end{figure*}

\subsection{Sample preparation}
\label{subsec:sample-prep}

The polymers were dissolved in dichloretane up to concentration 3
weight \% and spin coated on the quartz slides. The obtained films
were kept at room temperature over one day for evaporation of the
solvent. The thickness of the films of about 200-600~nm was measured
with a profilometer.  The UV/Vis spectra of the films measured by
silicon array spectrometer (OceanOptics Inc.) are presented in
Fig.~\ref{fig:UV-abs}. 
It can be seen that the position of the most intensive absorption
band ($\pi\pi^{*}$ band of azobenzene chromophore) depends on the end
substitute of the chromophore. The wavelength of the maximal
absorption, 
$\lambda_{max}$, for polymer P1 containing chromophore with push-pull
properties is shifted to the red as compared to 
$\lambda_{max}$ for P2.  

The anisotropy in the films was induced under irradiation
from two different sources of light with different wavelengths
$\lambda_{ex}$ and intensities $I$:
\begin{enumerate}
\item 
an Ar$^+$ laser with  $\lambda_{ex}= 488$~nm  and $I=0.7$~W/cm$^2$;

\item 
a mercury lamp with $\lambda_{ex}= 365$~nm and
$I=1.5$~mW/cm$^2$. In this case
the light was selected by an interference filter and polarized with a
Glan-Tompson polarization prism. 
\end{enumerate}
The wavelengths $365$~nm 
and $488$~nm  fall into $\pi\pi^{*}$ 
and n$\pi^{*}$
absorption band of azopolymers, respectively.  
In both cases the films were
irradiated at normal incidence of the actinic light. The irradiation
was carried out in several steps followed by studies of the
orientational structure. The resulting time of irradiation was
calculated by adding irradiation times of all irradiation steps
assuming accumulation of the structural changes. The time interval
between irradiation and structural studies was about 15~min. The
reason for time delay between these processes will be discussed below.

\subsection{Null ellipsometry method}
\label{subsec:null-ellips}

Instead of the prism coupling methods commonly used for the estimation 
of principal refractive indices we applied null ellipsometry 
technique~\cite{Azz:1977} dealing with birefringence components. 
By this means we have avoided 
some disadvantages of prism coupling method such as 
the problem of 
making optical contact between the prism and the polymer layer. 

\begin{figure*}[!tbh]
\centering
\resizebox{150mm}{!}{\includegraphics*{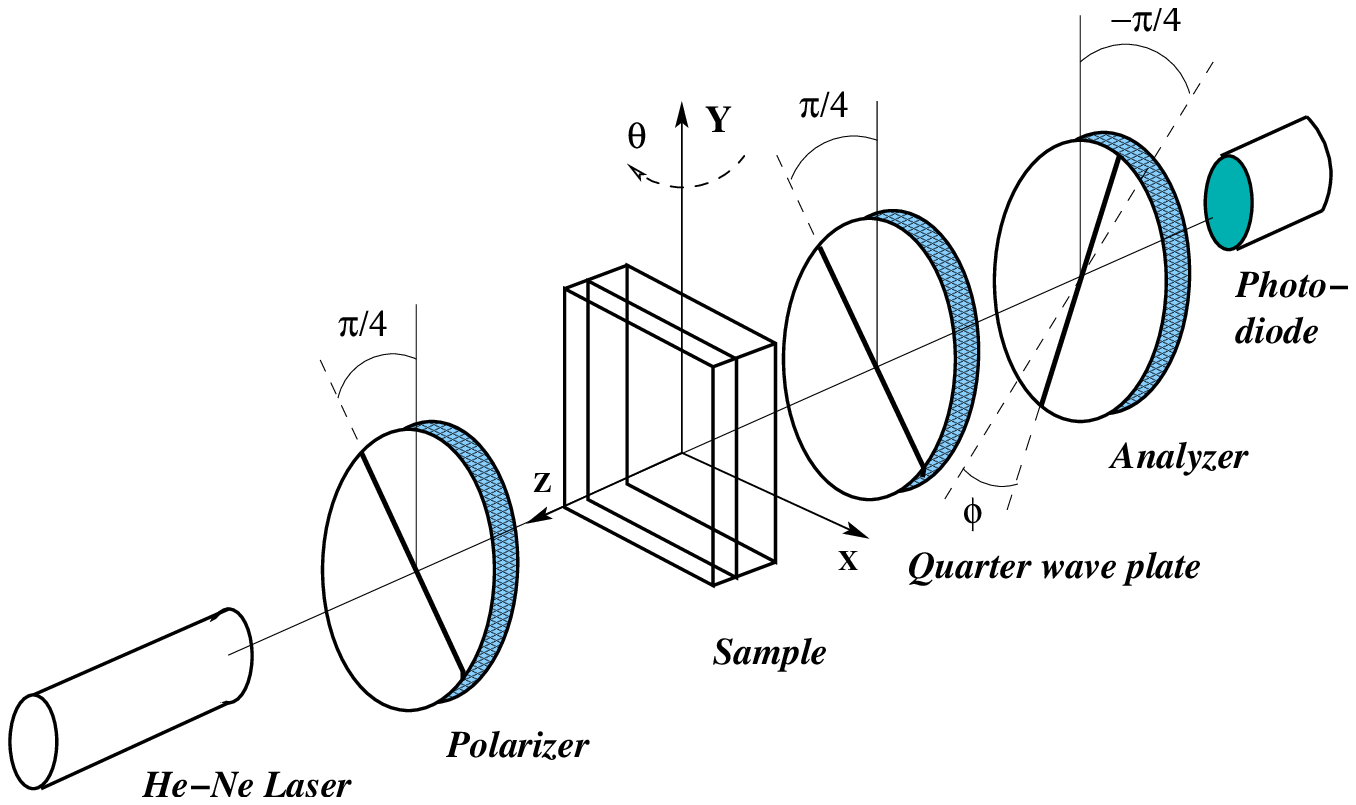}}
\caption{%
Null ellipsometry setup.  
}
\label{fig:setup}
\end{figure*}

The optical scheme of our method is presented in Fig.~\ref{fig:setup}.
The polymer film is placed between crossed polarizer and analyzer and
a quarter wave plate with the optic axes oriented parallel to the
polarization direction of the polarizer.  The light beam, which
wavelength is far away from the absorption band of the polymer, is
passing through the optical system.  The elliptically polarized beam
passing through the sample is transformed into the linearly polarized
light by means of the quarter wave plate.  The polarization plane of
this light is turned with respect to the polarization direction of the
polarizer.  This rotation is related to the phase retardation acquired
by the light beam after passing through the film under investigation.
It can be compensated by rotating the analyzer by the angle $\phi$
that encodes information on the phase retardation.
 
This method used for the normal incidence of the testing light is
known as the Senarmont technique.  It is suitable for the in-plane
birefringence measurements.

Using oblique incidence of the testing beam
we have extended this method for estimation of both in-plane, 
$n_y-n_x$, 
and out-of-plane $n_z-n_x$ 
birefringence ($n_x$, $n_y$ and $n_z$ are 
the principal refractive indices of the film). 
In this case, the angle $\phi$  depends on the 
in-plane retardation $(n_y-n_x)d$, the out-of-plane retardation 
$(n_z-n_x)d$ 
and the absolute value of a refractive index of the biaxial 
film, say, $n_x$.

\begin{figure*}[!thb]
\centering
\subfigure[non-irradiated P2 polymer with $n_z>n_y=n_x$]{%
\resizebox{75mm}{!}{\includegraphics*{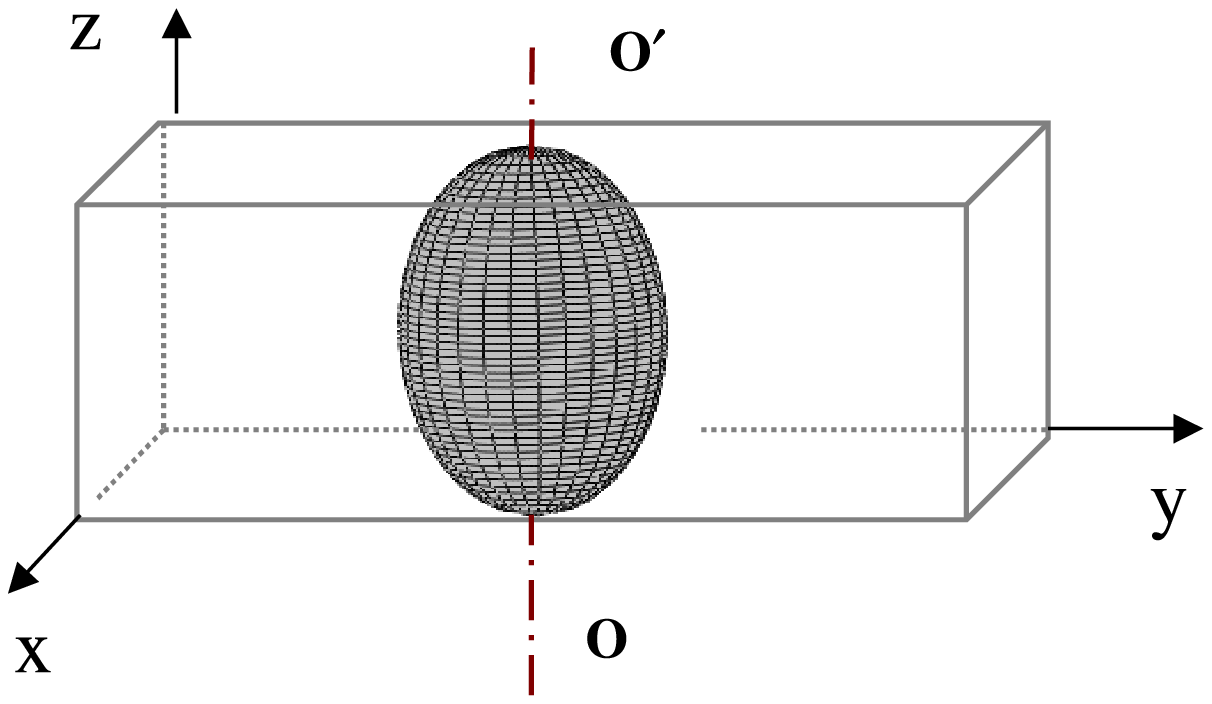}}
\label{fig:str-a}}
\subfigure[non-irradiated P1 polymer with $n_z<n_y=n_x$]{%
\resizebox{75mm}{!}{\includegraphics*{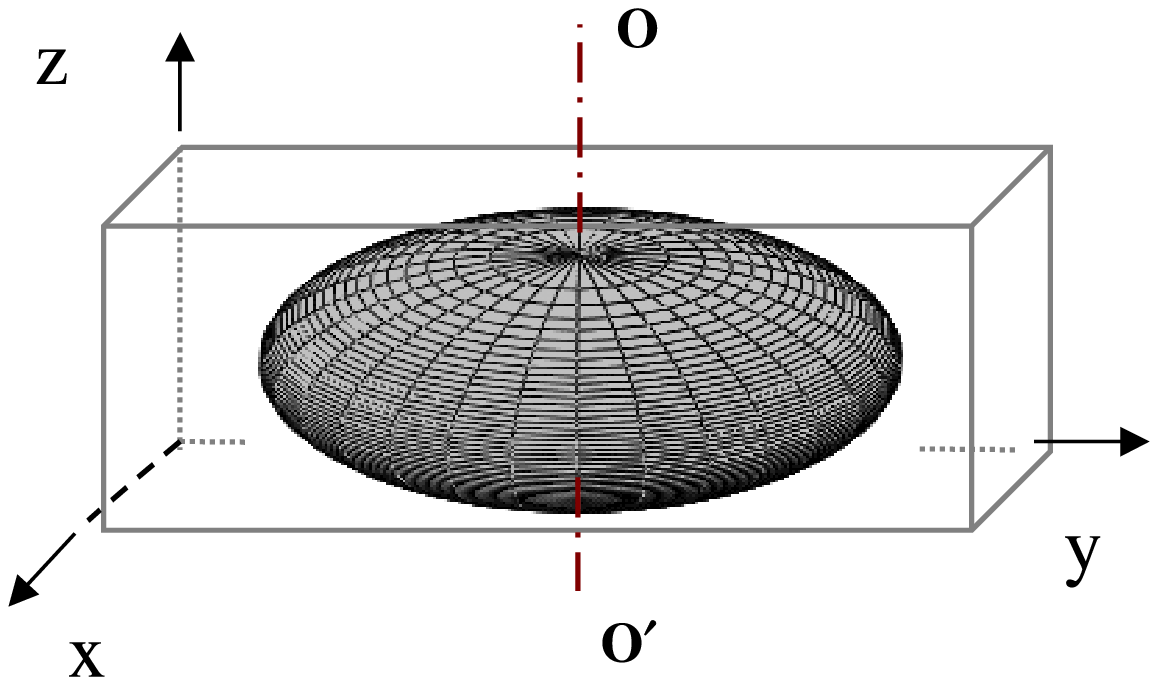}}
\label{fig:str-b}}
\subfigure[photo-saturated P1 polymer with $n_y>n_x=n_z$]{%
\resizebox{75mm}{!}{\includegraphics*{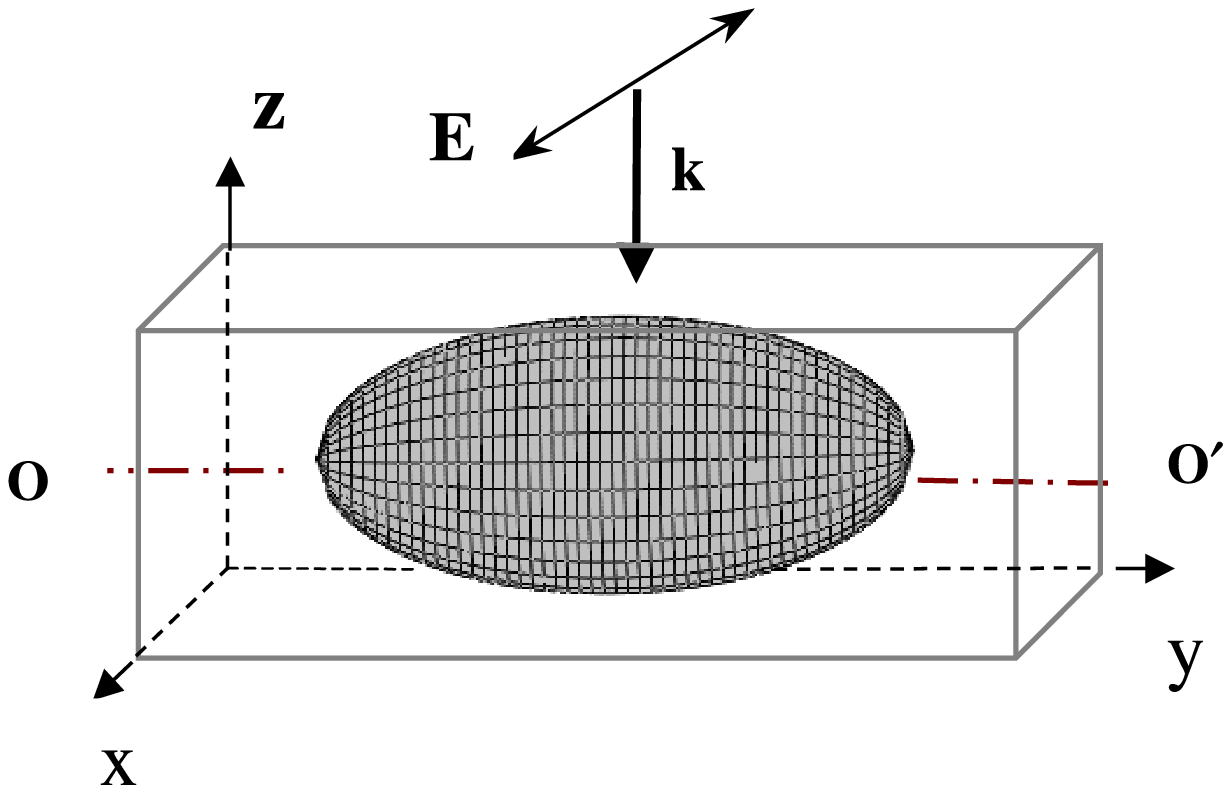}}
\label{fig:str-c}}
\subfigure[photo-saturated P2 polymer at $\lambda_{ex}=365$~nm
with $n_z=n_y=n_x$]{%
\resizebox{75mm}{!}{\includegraphics*{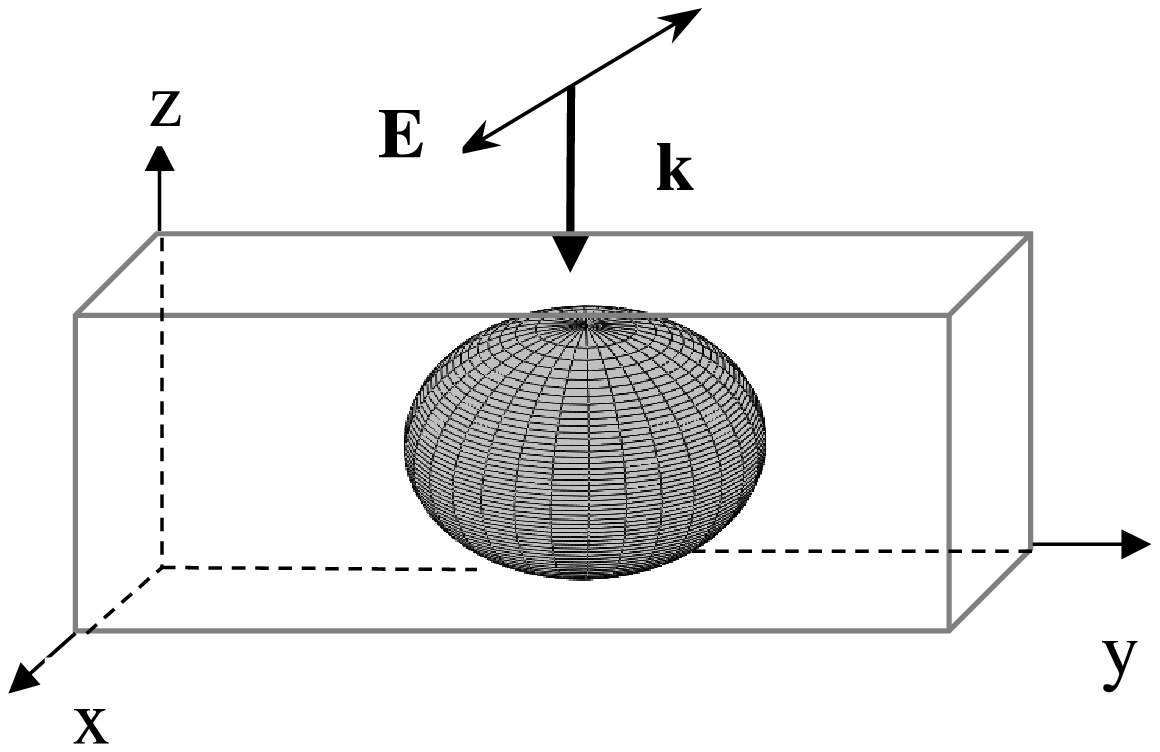}}
\label{fig:str-d}}
\subfigure[photo-saturated P2 polymer at $\lambda_{ex}=488$~nm
with $n_x<n_y=n_z$]{%
\resizebox{90mm}{!}{\includegraphics*{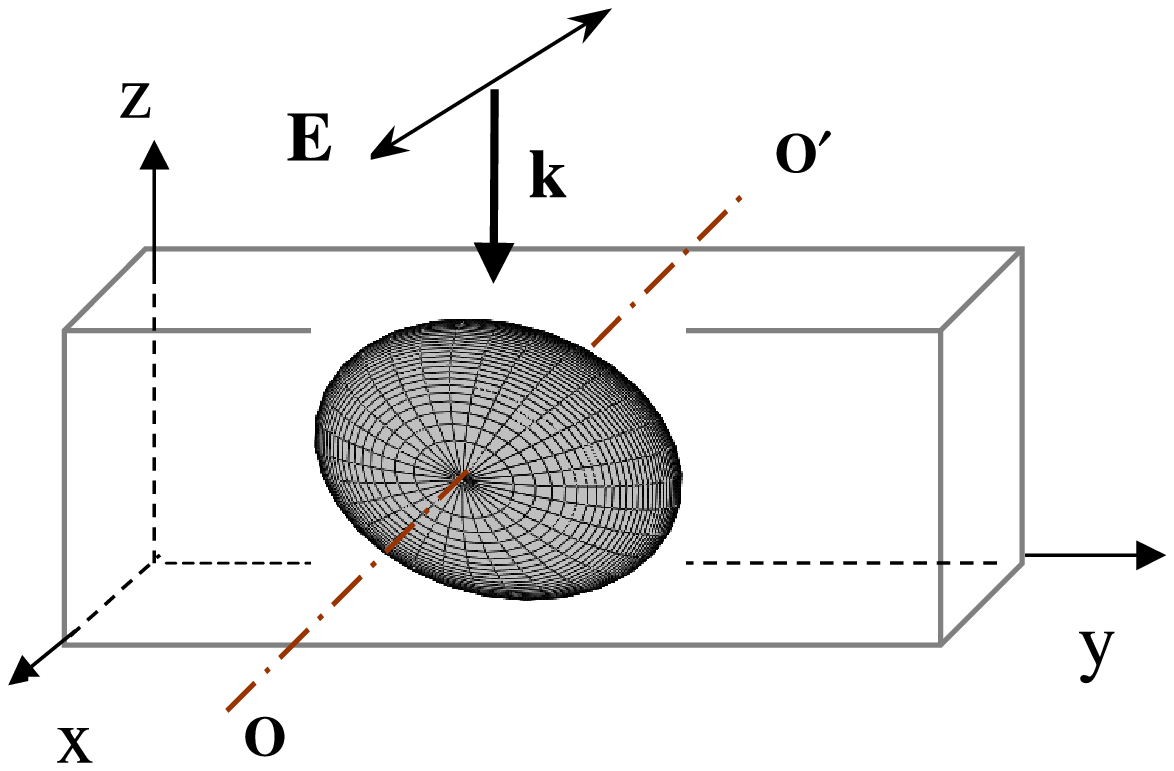}}
\label{fig:str-e}}
\caption{
Ellipsoids of refractive indices
for non-irradiated and photo-saturated polymer films.
}
\label{fig:str}
\end{figure*}

We need to have
the light coming out of the quarter wave plate  almost linearly 
polarized when the system analyzes the phase shift between two 
orthogonal eigenmodes of the sample. 
In our experimental setup this requirement 
can be met, when the $x$ axis, directed along the 
polarization vector of the actinic light, is oriented horizontally or 
vertically. 
Dependencies of the analyzer rotation angle $\phi$ 
on the incidence angle of the testing beam $\theta$   
were measured for both vertical and 
horizontal orientation of the $x$ axis.
The value of $n_x$ was 
measured  with the Abbe refractometer independently. 
 
By using Berreman's 
$4\times 4$ 
matrix method~\cite{Berr:1972},
the $\theta$-dependencies of $\phi$ were calculated. 
Maxwell's equations for the light propagation 
through the system of polarizer, sample and quarter wave plate were 
solved numerically for the different configurations of optical axes in 
the samples.
The measured  and computed $\phi$ versus $\theta$   
curves were fitted in the most probable configuration model 
using the measured value of $n_x$. 

We conclude on alignment of the azobenzene fragments 
from the obtained values 
of $(n_y-n_x)d$  and $(n_z-n_x)d$ assuming that the preferred direction of 
these fragments coincides with the direction of the largest refractive 
index. More details on the method can be found in our 
previous publication~\cite{Yar:2001}. 

In our setup designed for the
null ellipsometry measurements we used a low power He-Ne laser 
($\lambda=632.8$~nm), 
two Glan-Thompson polarizers mounted on rotational sta\-ges 
from Oriel Corp., a quarter wave plate from Edmund Scientific and a 
sample holder mounted on the rotational stage. The light intensity was 
measured with a photodiode. The setup was automatically controlled by 
a personal computer. 
The rotation accuracy of the analyzer was better than 
0.2~degree.

\subsection{Absorption method}
\label{subsec:abs-method}

The UV/Vis absorption measurements were carried out using a diode 
array spectrometer (OceanOptics). The samples were set normally 
to the testing beam of a deuterium lamp. 
A Glan-Thompson prism mounted on a 
computer-driven stepper was used to control 
the polarization of the testing beam, $\vc{E}_t$. 

The UV spectra of both original and irradiated films were 
measured in the spectral range from 250~nm to 600~nm 
for the probing light linearly polarized along the $x$ and $y$ axes.
From these data the optical density components
$D_x$  and $D_y$ were estimated at 
the absorption maximum of azochromophores,
$\lambda_t=375$~nm (for polymer P1) and $\lambda_t=343$~nm (for polymer P2).

In general, it is difficult to estimate the out-of-plane absorption
coefficient $D_z$. This, however, can be done in the regime of POA where
the fraction of \cis\ isomers is negligible
(the photo-reorientation mechanism) by using
the method proposed in Refs.~\cite{Wies:1992,Phaa:1996}.
The key point is that
the sum of all the principal absorption
coefficients, $D_{tot}=D_x+D_y+D_z$, does not depend on irradiation
doses.
So, if the anisotropy is known to be uniaxial at time $t_0$
with either 
$D_x(t_0)=D_z(t_0)$ or 
$D_y(t_0)=D_z(t_0)$, this
will yield the relation
\begin{align}
&
    D_{tot}=D_x+D_y+D_z\notag\\
&
=\begin{cases}
D_y(t_0)+2 D_x(t_0), &\vc{n}_o \parallel \uvc{e}_y,\\
D_x(t_0)+2 D_y(t_0), &\vc{n}_o \parallel \uvc{e}_x,\\
\end{cases}
\label{eq:D_tot}
\end{align}
where $\vc{n}_o$ is the optical axis
and $\uvc{e}_i$ is the $i$th coordinate unit vector. 
The out-of-plane component $D_z$ can now be computed
from Eq.~\eqref{eq:D_tot}.
In addition, we calculate the absorption
order parameters $S_i^{(a)}$ defined by the expression
\begin{equation}
  \label{eq:S_abs}
  S_i^{(a)}=\frac{2 D_i - D_j - D_k}{2(D_x+D_y+D_z)},\quad
i\ne j\ne k\,.
\end{equation}
In Sec.~\ref{sec:num-res} we find that
in the regime of photo-reorientation
the parameters $S_i^{(a)}$ are proportional to the diagonal components
of the order parameter tensor of \trans\ azochromophores.

\begin{figure*}[!tbh]
\centering
\resizebox{150mm}{!}{\includegraphics*{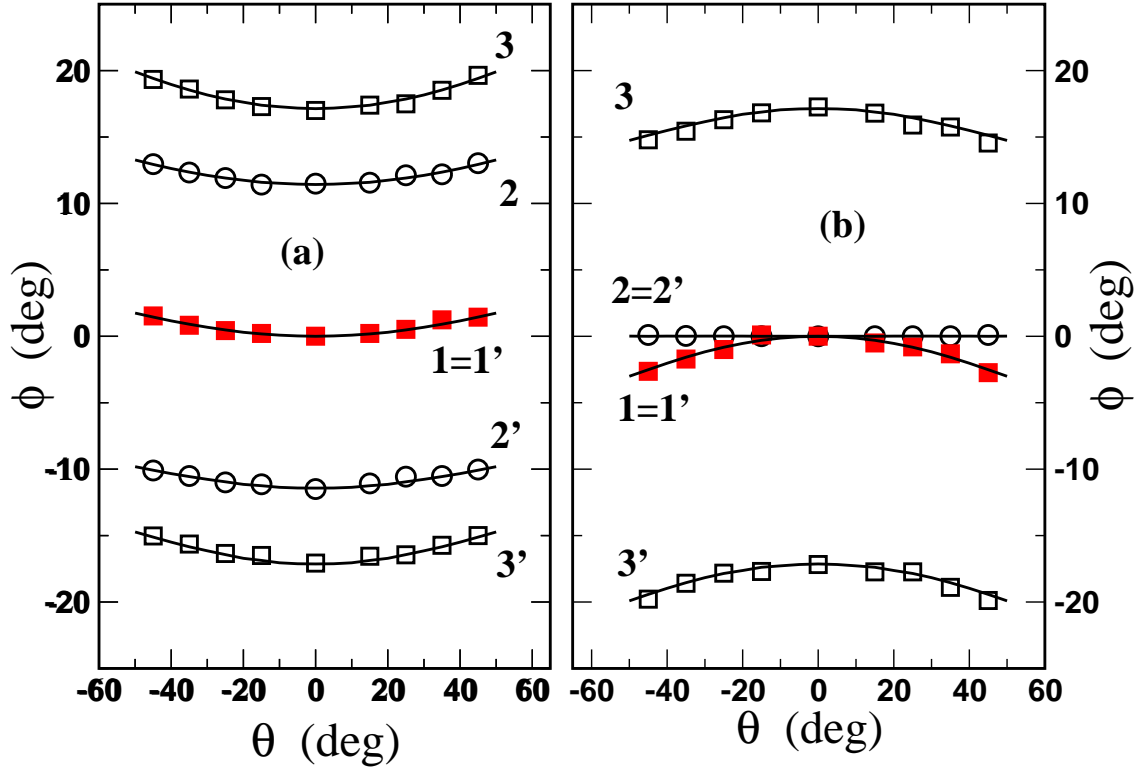}}
\caption{%
Measured (circles and squares) and modeled (solid lines) curves
for analyzer angle $\phi$ versus incidence angle $\theta$ 
for polymer P1 (a) and polymer P2 (b). 
}
\label{fig:phi-th}
\end{figure*}
  
\section{Experimental results}
\label{sec:exper-results}

\subsection{Non-irradiated films}
\label{subsec:non-irrad-films}

Fig.~\ref{fig:phi-th}(a) shows the experimentally measured curves of 
the phase
shift $\phi$ versus incidence angle $\theta$ for the polymer P1. The curve 1
corresponds to the non-irradiated film. The curves measured for
vertical and horizontal position of the $x$-axis overlap. This implies
that the in-plane principle indices $n_x$ and $n_y$ are matched. The film,
however, possess out-of-plane birefringence. The fitting gives the
value $(n_z-n_x)d=-20$~nm. Using the value of the film
thickness $d=200$~nm, we have $n_z-n_x=-0.1$ and 
the relation $n_x=n_y<n_z$ that
suggests preferred in-plane alignment with random orientation of
azobenzene fragments in the plane of the film. Optically, this structure
corresponds to the negative C plate depicted in Fig.~\ref{fig:str-a}.  

Fig.~\ref{fig:phi-th}(b) shows the measured $\phi$ versus $\theta$ curves 
for the polymer P2.  
The results obtained for the non-irradiated film (curve 1) imply that,
similar to the polymer P1, the in-plane birefringence is negligibly
small 
$\Delta n_{yx}= n_y-n_x\approx 0$. 
But the film is now
characterized by positive out-of-plane birefringence: $(n_z-n_x)d=35$~nm
($\Delta n_{zx}=n_z-n_x\approx 0.09$) and $n_z>n_x=n_y$. 
So, the film of the
polymer P2 is a positive uniaxial medium with the optical axis normal to
the film surface (positive C film shown in Fig.~\ref{fig:str-b}) and the
azobenzene fragments oriented homeotropically.

\begin{figure*}[!tbh]
\centering
\resizebox{120mm}{!}{\includegraphics*{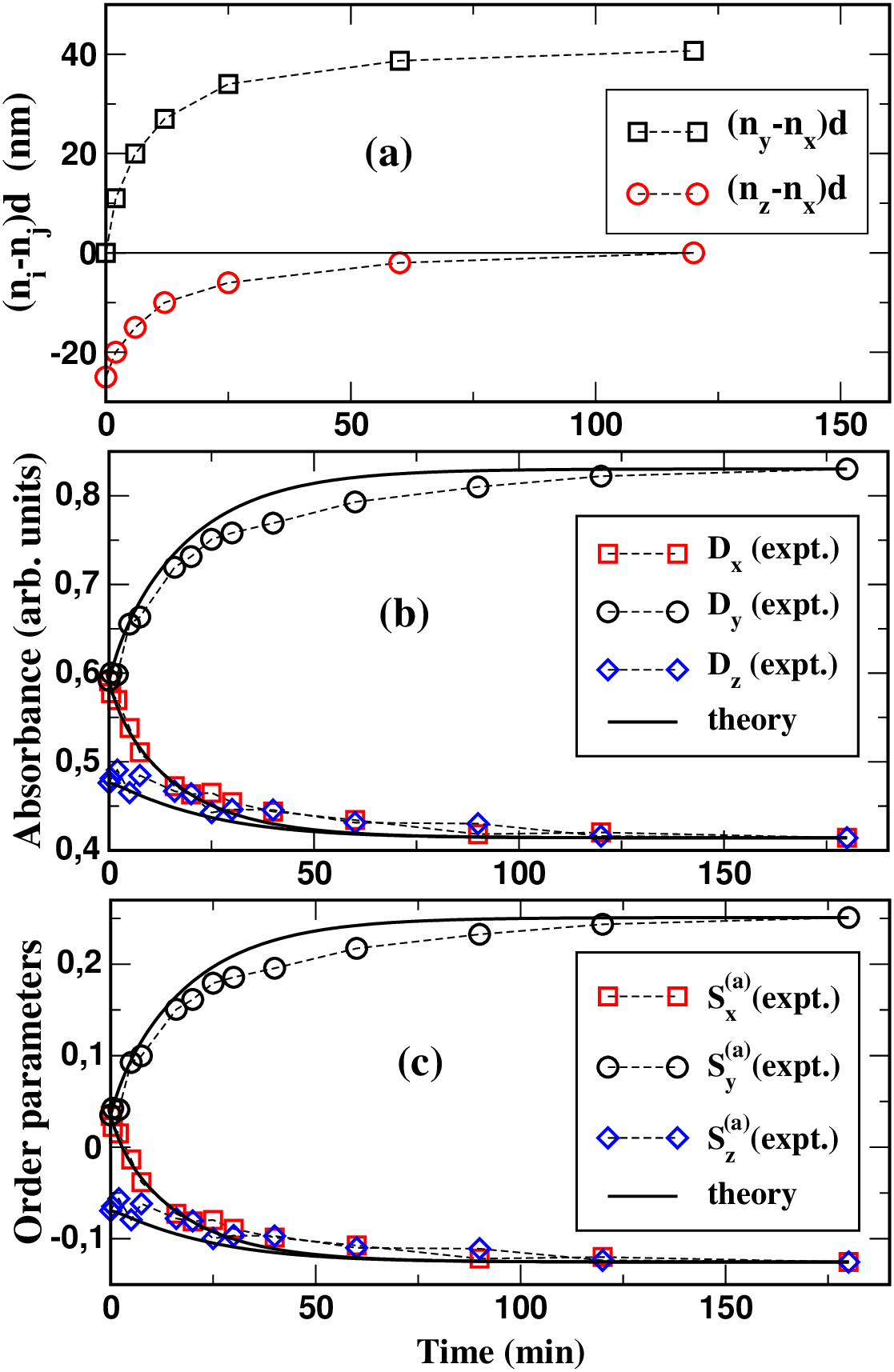}}
\caption{%
(a) Birefringence,
(b) principal absorption coefficients, and
(c) order parameter components  
as a function of irradiation time
at $\lambda_{ex}=365$~nm and $I=1.5$~mW/cm$^2$
for polymer P1.
}
\label{fig:P1-365}
\end{figure*}

\begin{figure*}[!tbh]
\centering
\resizebox{120mm}{!}{\includegraphics*{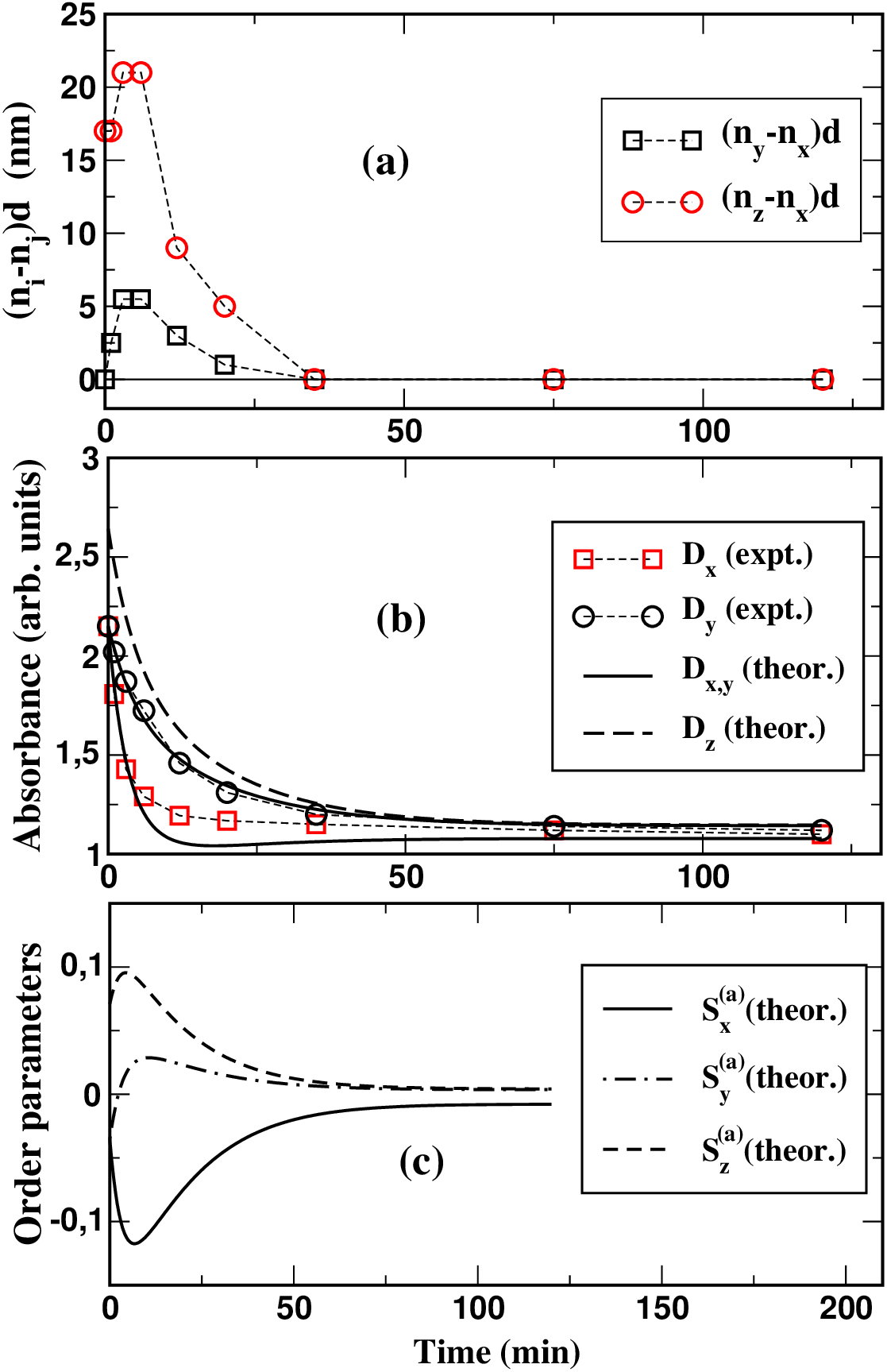}}
\caption{%
(a) Birefringence,
(b) principal absorption coefficients, and
(c) order parameter components  
as a function of irradiation time 
at $\lambda_{ex}=365$~nm and $I=1.5$~mW/cm$^2$
for polymer P2.
}
\label{fig:P2-365}
\end{figure*}

\subsection{Irradiation at $\lambda_{ex}= 365$ nm}
\label{subsec:irrad-with-365}

\subsubsection{Polymer P1}
\label{subsubsec:p1-365}

The curves 2 and 2' in Fig.~\ref{fig:phi-th}(a) show 
the measured $\phi$ versus $\theta$
curves for the polymer P1 after 60~min of UV light irradiation. Curves
2 and 2' correspond to vertical and horizontal position of $x$-axis of
the film. 
Since negative and positive phase shift at $\theta=0$ correspond to
the $x$ axis in the horizontal and vertical direction, respectively, the
higher in-plane refractive index is $n_y$ (i.e., in the direction
perpendicular to UV light polarization) and the lower one is $n_x$. 
Curve
fitting gives the following relation between the refractive
indices: $n_y-n_x=0.2$ ($(n_y-n_x)d\approx 40$~nm), 
$(n_z-n_x)d=0$~nm, $n_y>n_x=n_z$. So, the
photo-modified film is optically equivalent to the positive A plate
having optical axis in the plane of the film (Fig.~\ref{fig:str-c}). 
In this case, the
azobenzene fragments show planar alignment perpendicular to the UV
light polarization.  

The fitted values of the in-plane, 
$(n_y-n_x)d$, and
out-of plane, $(n_z-n_x)d$, retardation for P1 corresponding to various
irradiation times $\tau_{ex}$ are presented in 
Fig.~\ref{fig:P1-365}(a). 
The in-plane birefringence 
increases and saturates as the irradiation time increases.
Whereas the out-of-plane
birefringence is a decreasing function of the irradiation time 
and the difference between $n_z$ and 
$n_x$ becomes negligible in the saturation state. At early stages of
irradiation all the refractive indices are different  
$n_z<n_x<n_y$ and the film is biaxial. In the saturation
state the relation  $n_y>n_x=n_z$ implies that the ordering of azobenzene
chromophores is uniaxial.

Fig.~\ref{fig:P1-365}(b) represents the results of the
absorption measurements for P1 film before irradiation and after
subsequent irradiation steps. 
In accord with the null ellipsometry results
after irradiation $D_y$ is above $D_x$ and the azochromophores
are aligned along the $y$ 
direction that is perpendicular to $\vc{E}$.

The curves $D_x(\tau_{ex})$ and 
$D_y(\tau_{ex})$ are
typical of the photo-reorientation 
mechanism~\cite{Dum:1992,Dum:spie:1994}. 
In this case the fraction of
\cis\ isomers is negligible and the out-of-plane absorption
coefficient $D_z$ depicted in Fig.~\ref{fig:P1-365}(b) can be estimated 
by using Eq.~\eqref{eq:D_tot} to calculate $D_{tot}$
in the photo-saturated state where $D_x^{(st)}=D_z^{(st)}$.
Fig.~\ref{fig:P1-365}(c) shows the absorption
order parameters $S_i^{(a)}$ computed from the expression~\eqref{eq:S_abs}.

\subsubsection{Polymer P2}
\label{subsubsec:p2-365}

The curve 2 in Fig.~\ref{fig:phi-th}(b) shows the
measured $\phi$ versus $\theta$ function for the film of P2 irradiated with UV
light over 60~min. The curves corresponding to the vertical and the horizontal
positions of the $x$ axis of the film overlap. 
The fitting gives $n_x=n_y=n_z$ and, thus,
the angular distribution of azobenzene
chromophores is isotropic.  

Fig.~\ref{fig:P2-365}(a) shows the fitted values of $(n_y-n_x)d$ and
$(n_z-n_x)d$, obtained for various irradiation times of P2. The
dependencies of $(n_y-n_x)d$ and $(n_z-n_x)d$ on the irradiation time
go through 
the maximum and saturate at large irradiation doses. At the initial
irradiation stage $n_z>n_y>n_x$ and the film is biaxial. 
In the photosaturated state
$n_x=n_y=n_z$ and, as is illustrated in Fig.~\ref{fig:str-d}, 
the film is isotropic.  

The
experimentally measured 
$D_x(\tau_{ex})$ and $D_y(\tau_{ex})$ curves are shown in 
Fig.~\ref{fig:P2-365}(b). 
Both curves decrease with increasing the irradiation dose.
This behavior is typical of the mechanism of
photoselection~\cite{Dum:1992,Dum:spie:1994}.
In this regime the exciting light 
will cause angular selective depopulating of
the \trans\ state, so that the fraction
of \trans\ isomers rapidly grows smaller with illumination time.
This 
process~--~the so-called angular selective burning~--~dominates 
the kinetics of POA and
gives rise to diminution of both absorption coefficients 
$D_x$ and $D_y$. 
The isotropy of the photosaturated state then can be
explained by negligibly small concentrations of \trans\ chromophores
at large irradiation doses. This point will be discussed at greater
length in Sec.~\ref{sec:discussion}.

\begin{figure*}[!tbh]
\centering
\resizebox{120mm}{!}{\includegraphics*{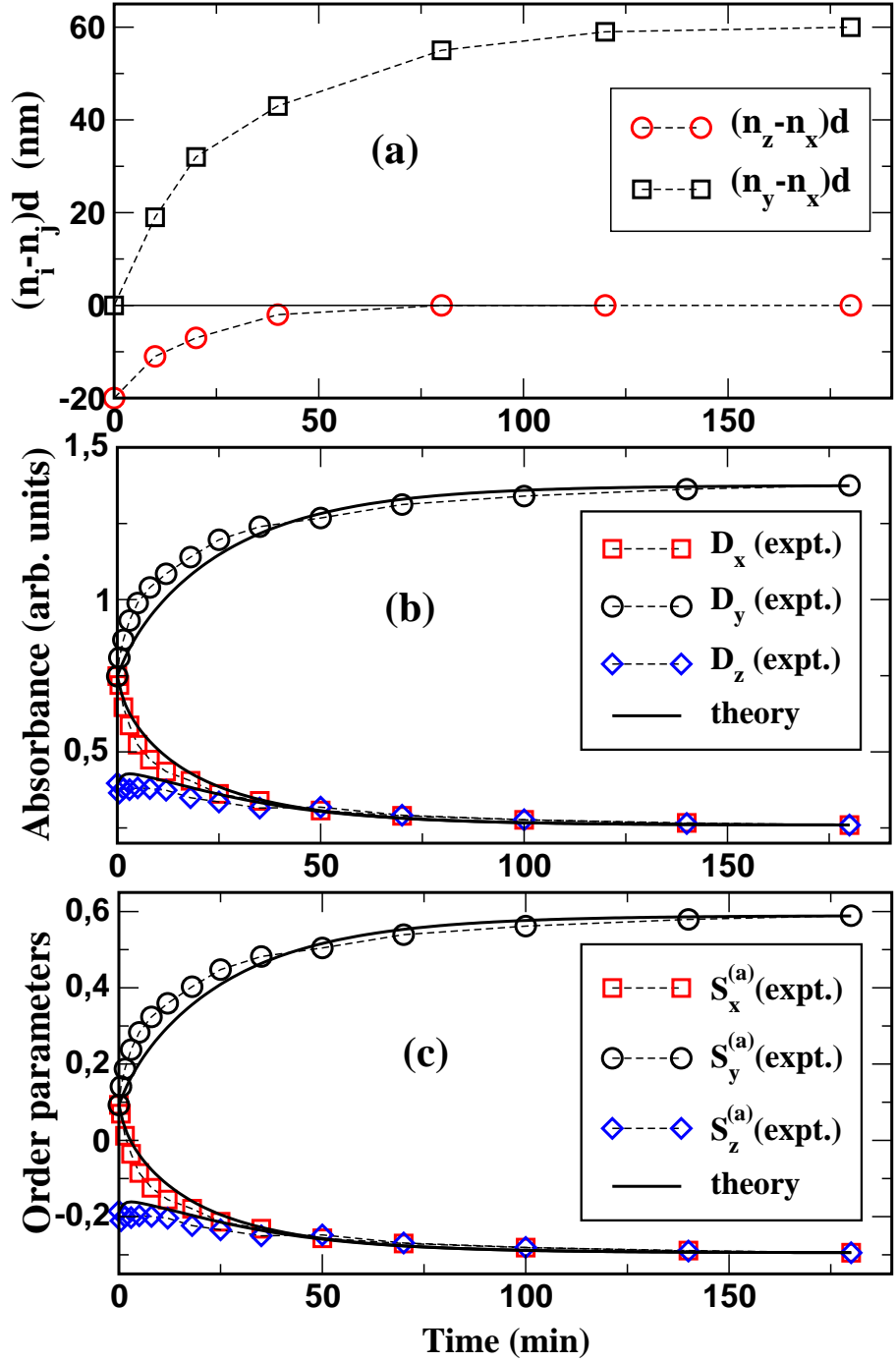}}
\caption{%
(a) Birefringence,
(b) principal absorption coefficients, and
(c) order parameter components  
as a function of irradiation time 
at $\lambda_{ex}=488$~nm and $I=0.7$~W/cm$^2$
for polymer P1.
}
\label{fig:P1-488}
\end{figure*}

\begin{figure*}[!tbh]
\centering
\resizebox{120mm}{!}{\includegraphics*{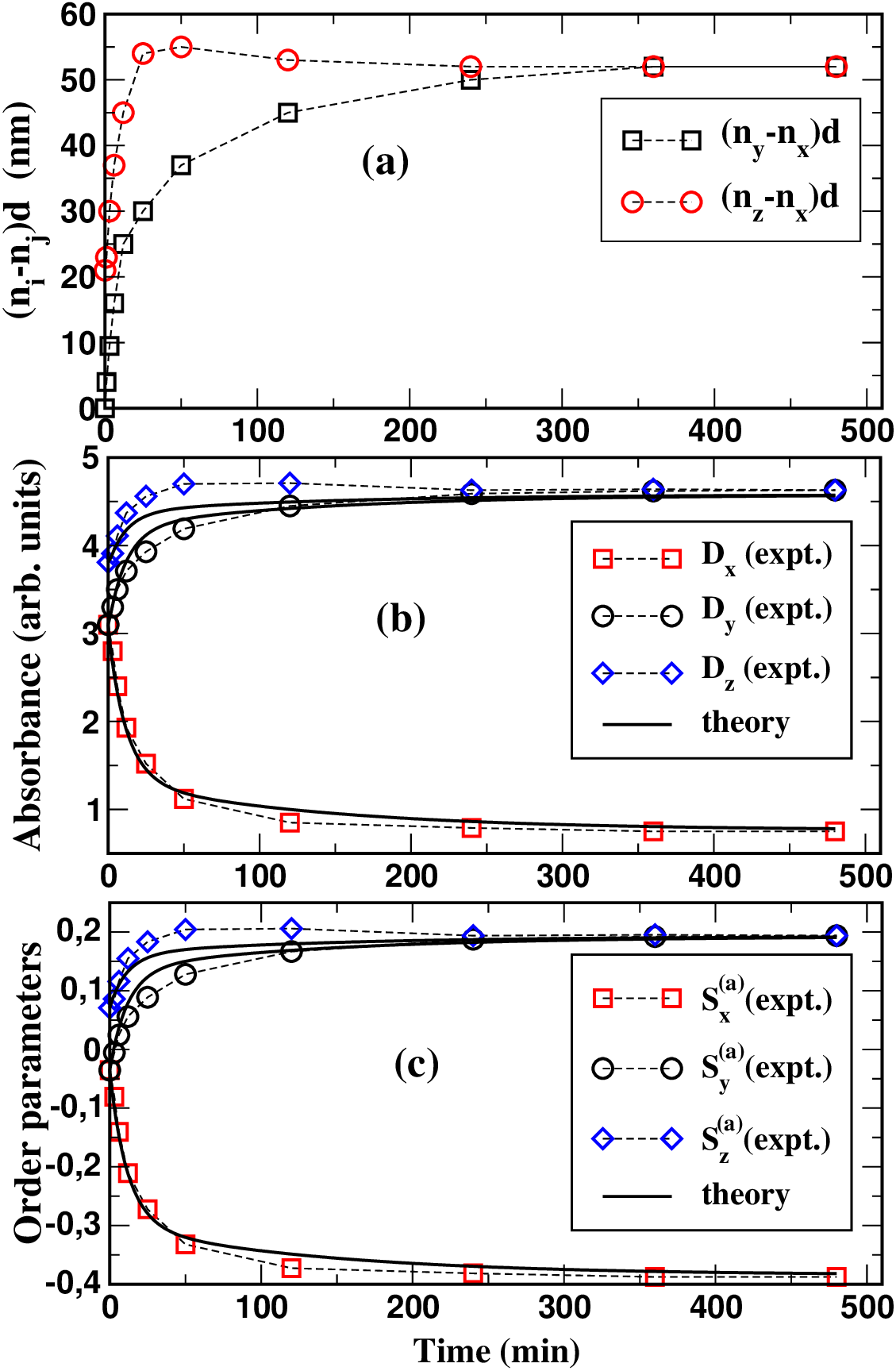}}
\caption{%
(a) Birefringence,
(b) principal absorption coefficients, and
(c) order parameter components  
as a function of irradiation time 
at $\lambda_{ex}=488$~nm and $I=0.7$~W/cm$^2$
for polymer P2.
}
\label{fig:P2-488}
\end{figure*}

\subsection{Irradiation at $\lambda_{ex}=488$~nm}
\label{subsec:irrad-with-488}

\subsubsection{Polymer P1}
\label{subsubsec:p1-488}

In Fig.~\ref{fig:phi-th}(a) the curves corresponding to irradiation with 
$\lambda_{ex}=488$ nm ($\tau_{ex}=300$ min) are denoted as 3 and 3'. 
Similar to the case where $\lambda_{ex}=365$~nm, 
the relation between the principal refractive indices
is $n_y>n_x=n_z$. 
So, the induced orientational
configurations do not differ and correspond to 
a uniaxial in-plane alignment with the optical 
properties of a positive A plate.  
From the other hand, 
the value of in plane retardation $(n_y-n_x)d$ is
considerably higher than it is for $\lambda_{ex}=365$~nm and  
reaches 60. 

The $(n_y-n_x)d$ and 
$(n_y-n_x)d$ versus $\tau_{ex}$ kinetic curves
for the film and irradiation conditions 
under consideration are presented in
Fig.~\ref{fig:P1-488}(a). 
The results are qualitatively similar to those obtained
for $\lambda_{ex}=365$~nm: 
biaxial alignment of the azobenzene chromophores at the early stages
of irradiation and  
the uniaxial structure in the photosaturated state.  
The $D_x(\tau_{ex})$ and $D_y(\tau_{ex})$ curves
presented in Fig.~\ref{fig:P1-488}(b) 
indicate in-plane angular redistribution of
the azobenzene chromophores and the preferential alignment along the
direction perpendicular to $\vc{E}$.
In addition, we used Eq.~\eqref{eq:D_tot} to compute 
the curve for the out-of-plane component $D_z(\tau_{ex})$
plotted in Fig.~\ref{fig:P1-488}(b). 
Fig.~\ref{fig:P1-488}(b) shows dependencies of
the absorption order parameters on the irradiation time
calculated from Eq.~\eqref{eq:S_abs}. 

\subsubsection{Polymer P2}
\label{subsubsec:p2-488}

The curves 3 and 3' in Fig.~\ref{fig:phi-th}(b)
show the measured $\phi$ versus $\theta$ dependencies for the film of P2
irradiated with the actinic light at $\lambda_{ex}= 488$ nm over 300~min. 
As above, 
the curves corresponding to the vertical and the horizontal position of
the $x$ axis are labelled 3 and 3', respectively. The fitting yields 
$(n_y-n_x)d=(n_z-n_x)d=60$~nm
and $n_y=n_z>n_x$. So, the induced orientational
structure is uniaxial. 
The optical axis of this structure lies in the
film plane and is directed along the polarization vector of the actinic
light. 
The principal refractive index for the optical axis direction has the
lowest value. The same optical properties possess
crystal plate called as negative A plate (Fig.~\ref{fig:str-e}). 
The negative A plate
in case of azopolymer film stands for random orientation of
azochromophores in the plane perpendicular to vector 
$\vc{E}$. 
This is precisely the structure which can be expected from
the na\"{\i}ve symmetry considerations discussed 
in Sec.~\ref{sec:intro}.  

The fitting results,
$(n_y-n_x)d$ and $(n_y-n_x)d$, plotted as functions of 
irradiation time $\tau_{ex}$
are presented in Fig.~\ref{fig:P2-488}(a). The curves converge approaching 
the saturated state. The transient photoinduced structures are
biaxial. 

The $D_x(\tau_{ex})$ and $D_y(\tau_{ex})$ curves  
for the case under investigation are shown in
Fig.~\ref{fig:P2-488}(b).
By contrast to the case presented in Fig.~\ref{fig:P2-365}(b),
these curves clearly indicate that the regime of POA
is dominated by  the photo-reorientation
mechanism~\cite{Dum:1992,Dum:spie:1994}.
So, the out-of-plane coefficient $D_z$ can be estimated
by applying the procedure described in Sec.~\ref{subsec:abs-method}
to the case in which the known uniaxial structure
is represented by the state of saturation with $D_y^{(st)}=D_z^{(st)}$.
The curve $D_z(\tau_{ex})$ computed from 
Eq.~\eqref{eq:D_tot} and dependencies of the
absorption order parameters on the irradiation time
are depicted in Figs.~\ref{fig:P2-488}(b) 
and~\ref{fig:P2-488}(c), respectively.

\begin{figure*}[!tbh]
\centering
\resizebox{130mm}{!}{\includegraphics*{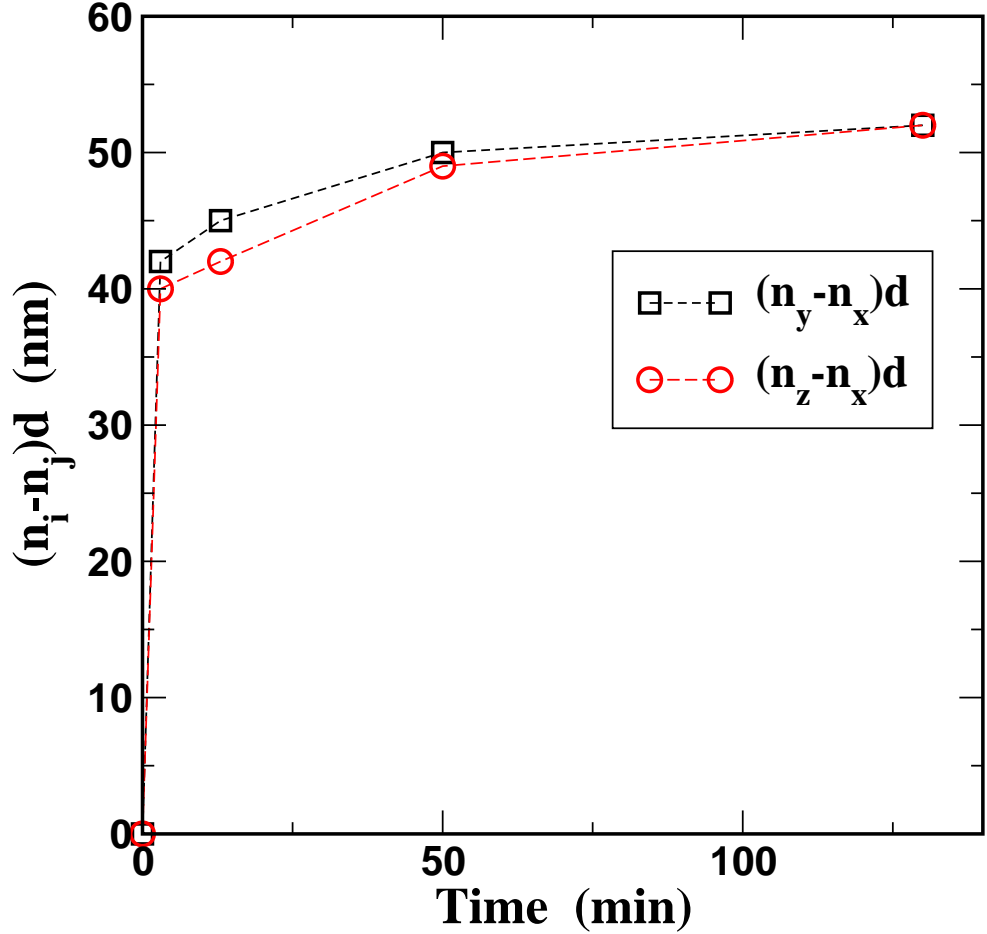}}
\caption{%
Birefringence
versus irradiation time
for the pre-irradiated polymer P2
(see discussion in Sec.~\ref{sec:subs-irrad-with}).
}
\label{fig:two-stp}
\end{figure*}

\subsection{Two step irradiation at $\lambda_{ex}= 365$~nm 
and $\lambda_{ex}= 488$ nm}
\label{sec:subs-irrad-with}

In these experiments 
the polymer films were initially irradiated with non-polarized
UV light ($\lambda_{ex}=365$~nm, 
$I=15$~mW/cm$^2$) for about an hour
and, subsequently, with the polarized light
from Ar$^+$ laser ($I=0.7$~W/cm$^2$). 
The first step of
irradiation does not produce considerable changes in the initial structure
of the polymer P1. The
$(n_y-n_x)d$ and $(n_y-n_x)d$ versus $\tau_{ex}$ 
kinetic curves obtained for the
second step of irradiation are very close to those obtained for
irradiation with only $\lambda_{ex}=488$~nm 
(see Fig.~\ref{fig:P1-488}(a)).  

After the first step of UV irradiation  
P2 film was optically isotropic.
But similar to the case of one step irradiation
at $\lambda_{ex}= 488$~nm, 
the subsequent irradiation have generated 
the anisotropy of a negative A film.
The corresponding $(n_y-n_x)d$ and $(n_y-n_x)d$ versus $\tau_{ex}$
kinetic curves are presented in Fig.~\ref{fig:two-stp}.  
It is seen that pre-irradiation with UV light strongly
accelerate formation of the saturated structure
as compared to the case of one step
irradiation (see Fig.~\ref{fig:P2-488}(a)). 
But for both P1 and P2 films 
this structure in itself is not affected by pre-irradiation.
Thus, the structure is completely determined by the
last step of irradiation.

\subsection{Thermal stability of the photoinduced structures}
\label{sec:therm-stab-induc}

At room temperature, we have not detected any changes in 
ordering of non-irradiated films for several weeks. 
The photoinduced uniaxial and biaxial
anisotropies, formed
in polymer films after switching off the exciting light,
persisted over the same period of time.

We have also studied the influence of the elevated temperatures on the
orientational structures. The baking of the non-irradiated P1
films at temperatures of mesophases and isotropic phase did not
change considerably the initial order. In contrast, P2 films had the
out-of-plane alignment of azobenzene chromophores enhanced after the
treatment at temperatures of both nematic mesophase and isotropic
phase.  For instance, baking of P2 film at 150\degc\ over 5~min led to
the increase in birefringence from $(n_e-n_o)d=35$~nm to
$(n_e-n_o)d=55$~nm.

The curing of P1 films at the temperatures of nematic and smectic
mesophases followed by cooling down to room temperature destroys
biaxiality of the photoinduced structures and enhances the uniaxial
anisotropy.  Similar effects were observed at the short time 
(about 5~min) curing in the isotropic state at 70\degc. 
In this case the prolonged curing caused the disappearance 
of the induced anisotropy.

The curing of P2 films at temperatures of nematic and isotropic phases
transformed the structure of negative uniaxial medium into 
the positive uniaxial configuration with 
azochromophores aligned perpendicular to
the film substrate.  

\subsection{Photochemical parameters}
\label{subsec:phot-param}
  
There are a number of parameters used in subsequent calculations
for theoretical interpretation of the experimental results.
In this subsection we dwell briefly on the experimental
results used to estimate
the lifetime of the \cis\ form, $\tau_c$, and
the absorption cross sections of isomers, $\sigma^{(cis)}$ and
$\sigma^{(tr)}$.  

\subsubsection{Lifetime of cis isomers}

The lifetime $\tau_c$ of 
\cis\ isomers was estimated by studying 
the relaxation of UV spectrum modified by irradiation with 
non-polarized light at $\lambda_{ex}= 365$~nm. The
pumping and testing light were directed almost normally to the
film. The wavelength of non-polarized testing light was adjusted at
the maximum of $\pi\pi^{*}$ absorption band of the studied polymer 
(375~nm and 343~nm for polymers P1 and P2, respectively).

Temporal evolution of the absorbance after switching off the pumping
was measured. 
The curves then were fitted 
by using the two-exponential approximation taken in the following form
\begin{equation}
  \label{eq:fit-uv}
  y(t)=y_0+A_1{\mathrm e}^{-t/\tau_1}+A_2{\mathrm e}^{-t/\tau_2}.
\end{equation}
For  polymer P1 we have 
$\tau_1=1.2$~s and
$\tau_2=1.3$~min with $A_1/A_2\approx 5.3$.
The relaxation curve obtained for polymer P2 is fitted well 
by Eq.~\eqref{eq:fit-uv} at
$\tau_1=2.6$~min,
$\tau_2=440$~min and $A_1/A_2\approx 0.13$.

In the case of polymer P1 relaxation is dominated by the fast
component that corresponds to the largest amplitude and can be
attributed to \cis-\trans\ backward isomerization of the chromophores
trapped in a strained conformation~\cite{Paik:1972}. The lifetime can
now be estimated as a decay time for the bulk of \cis\ isomers to
yield $\tau_c\approx 1.2$~s.

The origin of slower and less pronounced component could be caused by
thermal \cis-\trans\ isomerization of the other fraction of azobenzene
moieties, which are strongly restricted in their molecular dynamics by
the polymer matrix. Their reaction requires a stronger reorganization
of the polymer film~\cite{Eisen:1980}.  Since the waiting time in our
experiments was about 15~min, we can safely assume all \cis\ isomers
in polymer P1 relaxed back to the \trans\ isomeric form.

For polymer P2 the largest contribution comes from the slowest
component with $\tau_2=440$~min. This time gives an estimate for the
lifetime of \cis\ isomers, $\tau_c\approx 440$~min.

There is a difficulty in estimating $\tau_c$ with this
method.  In addition to \trans-\cis\ photoisomerization,
the UV spectrum can also be influenced by photo-reorientation.  
The non-polarized light
may cause out-of-plane reorientation of azobenzene chromophores
reducing the UV absorption. This
out-of-plane reorientation, however, cannot be effective in polymer P1
which shows strong preference to in-plane reorientation.  For 
polymer P2 the out-of-plane reorientation of azochromophores has been
additionally checked by the absorption measurement at the oblique
incidence of testing beam. An increase in the polymer absorption at
343~nm under UV irradiation has not been detected.  
Thus it is safe to assume that the observed relaxation can be
solely attributed to \cis-\trans\ thermal isomerization.

\begin{table*}[htbp]
\begin{ruledtabular}
  \begin{tabular}{cD{.}{.}{-1}D{.}{.}{-1}|cD{.}{.}{-1}D{.}{.}{-1}}
\multicolumn{3}{c|}{Polymer P1}&
\multicolumn{3}{c}{Polymer P2}\\
\colrule 
$\lambda_{ex}$ (nm) &365 & 488 &
$\lambda_{ex}$ (nm) &365 & 488 
\\
\colrule
$I$ (mW/cm$^2$) & 1.5 & 700 &
$I$ (mW/cm$^2$) & 1.5 & 700
\\
$\sigma_{cis}$ $\times 10^{-19}$(cm$^2$) & 24.8 & 0.8 &
$\sigma_{cis}$ $\times 10^{-19}$(cm$^2$) & 6.0 & 1.1
\\
$\sigma^{(tr)}/\sigma_{cis} $ & 5.6 & 0.47 &
$\sigma^{(tr)}/\sigma_{cis} $ & 57.0 & 0.2
\\
$\sigma_{||}^{(tr)}/\sigma_{\perp}^{(tr)}$ & 6.95 & 36.2 &
$\sigma_{||}^{(tr)}/\sigma_{\perp}^{(tr)}$ & 9.4 & 39.6
\\
$\Phi_{cis\to tr}$ (\%) & 10 & 10 &
$\Phi_{cis\to tr}$ (\%) & 5 & 10
\\
$\Phi_{tr\to cis}$ (\%) & 10 & 10 &
$\Phi_{tr\to cis}$ (\%) & 10 & 10
\\
$(\gamma_c+q_c I)/(q_t I)$  & 623.4 & 2181 &
$(\gamma_c+q_c I)/(q_t I)$  & 0.03 & 69.1
\\
  \end{tabular}
\end{ruledtabular}
  \caption{Photochemical parameters}
  \label{tab:param-pol}
\end{table*}

\subsubsection{Absorption cross sections}

The above mentioned photo-induced reorientation accompanying
photochemical process makes it extremely difficult to estimate the
coefficients of the molecular extinction in polymer films. For this
purpose we applied the method well known for polymer
solutions~\cite{Bern:1975}.  
Both polymers were dissolved in toluene at concentration 5 $10^{-3}$ g/l. The
UV/Vis spectra of the polymer solutions were measured before and
during irradiation. In the latter case the solutions were in the
photo-saturated state. 
Each spectrum was decomposed into three components of
Gaussian shape denoted as 
$D_c^{(i)}(\lambda)$, $D_t^{(i)}(\lambda)$ and 
$D_n^{(i)}(\lambda)$ in the expression
\begin{align}
D^{(i)}(\lambda)=
D_c^{(i)}(\lambda) + D_t^{(i)}(\lambda)+
D_n^{(i)}(\lambda),
\label{eq:D-sol}  
\end{align}
where the index $i$ labels the spectra of the non-irradiated 
($i=0$) and the irradiated solution ($i=1$).
The first two terms on the right hand side of Eq.~\eqref{eq:D-sol},
$D_c^{(i)}(\lambda)$ and $D_t^{(i)}(\lambda)$, 
can be assigned to $\pi\pi^{*}$ absorption of \cis\  
and \trans\ isomers, respectively. 
The last term, $D_n^{(i)}(\lambda)$, corresponds
to n$\pi^{*}$ absorption where the contributions from 
\cis\  and \trans\ isomers cannot be separated. 

We can now relate
the components $D_{t}^{(i)}$ 
and $D_{c}^{(i)}$ to the $\pi\pi^{*}$ 
absorption cross sections of isomers, 
$\sigma_{\pi}^{(cis)}$ and  $\sigma_{\pi}^{(tr)}$, 
and the concentrations of
isomers, $C_{tr}^{(i)}$ and $C_{cis}^{(i)}$,
in the non-irradiated film ($i=0$) and 
in the photo-saturated ($i=1$) state  as follows
\begin{subequations}
\label{eq:extin-pi}
\begin{align}
&
D_{t}^{(i)}(\lambda)=d C_{tr}^{(i)}\sigma_{\pi}^{(tr)}(\lambda) \, ,\\
&
D_{c}^{(i)}(\lambda)=d C_{cis}^{(i)} \sigma_{\pi}^{(cis)}(\lambda) \, ,\\
&
C=C_{tr}^{(i)} + C_{cis}^{(i)},\quad i=0,\,1,
\end{align}
\end{subequations}
where 
$d$ is the cell thickness,  $C$ is 
the experimentally measured total concentration of the
dissolved molecules.
Solving the system~\eqref{eq:extin-pi} yields the cross sections
$\sigma_{\pi}^{(\alpha)}$
and the concentrations $C_{\alpha}^{(i)}$ ($\alpha=tr,\,cis$)
which can be further employed to determine the
 absorption cross sections $\sigma_{n}^{(\alpha)}$
in the overlapping n$\pi^{*}$ absorption bands of isomers
from the relation
\begin{equation}
  \label{eq:extin-n}
 D_{n}^{(i)}=\bigl(\sigma_n^{(tr)} C_{tr}^{(i)} +
\sigma_n^{(cis)} C_{cis}^{(i)} \bigr) d.
\end{equation}

The average absorption cross section of \trans\
isomers $\sigma^{(tr)}$ and the cross section of \cis\
isomers $\sigma^{(cis)}$ at the wavelength $\lambda$
can now be computed as a sum of
$\sigma_{\pi}^{(\alpha)}(\lambda)$
and $\sigma_{n}^{(\alpha)}(\lambda)$.
The resulting estimates for the absorption cross sections of
the polymers calculated at the wavelengths of the
exciting light are given in Table~\ref{tab:param-pol}.
  
\section{Model}
\label{sec:model}

Theoretical considerations of this section deal with
the kinetics of POA that determines how amount of photoinduced anisotropy
characterized by either absorption dichroism or birefringence evolves
in time upon illumination and after switching it off.
Below we briefly discuss the simple phenomenological model
describing the kinetics of POA in terms of the order parameter tensor and
the concentrations of azochromophores. (More details on the model
and the underlying phenomenological approach can be found in 
Refs.~\cite{Kis:epj:2001,Kis:jpcm:2002}). 

We shall assume that the azobenzene groups in 
the ground state  are of \trans\  form (\trans\ molecules) 
and the orientation of the molecular axis is
defined by the unit vector 
$\uvc{n}=$($\sin\theta\cos\phi$,$\sin\theta\sin\phi$,
$\cos\theta$), where 
$\theta$ and $\phi$ are polar and azimuth angles of the unit vector.
Angular distribution of the \trans\  molecules at time $t$
is characterized by the number distribution function $N_{tr}(\uvc{n},t)$.
Similarly, azochromophores 
in the excited state have the \cis\  conformation (\cis\ molecules) and
are characterized by the function $N_{cis}(\uvc{n},t)$.
Then the number of \trans\  and \cis\  molecules is given by
\begin{gather}
  N_{tr}(t)\equiv N n_{tr}(t)=\int N_{tr}(\uvc{n},t)\,\dd\uvc{n},
  \label{eq:1t}\\
  N_{cis}(t)\equiv N n_{cis}(t)=\int N_{cis}(\uvc{n},t)\,\dd\uvc{n},
  \label{eq:1c}\\
n_{tr}(t)+n_{cis}(t)=1,
  \label{eq:consv}
\end{gather}
where 
$\displaystyle
\int\,\dd\uvc{n}\equiv 
\int_{0}^{2\pi}\dd\phi\int_{0}^{\pi}\,\sin\theta\,\dd\theta\,$ 
and 
$N$ is the total number of molecules.
The normalized angular
distribution functions, $f_\alpha(\uvc{n},t)$, of
\trans\  ($\alpha=tr$) and \cis\  ($\alpha=cis$) molecules 
can be conveniently defined by the relation
\begin{equation}
  \label{eq:dis}
  N_\alpha(\uvc{n},t)=N n_\alpha(t) f_\alpha(\uvc{n},t)\, . 
\end{equation}

We also need to introduce additional angular distribution function
$f_p(\uvc{n},t)$ characterizing the anisotropic field due to
interaction between a side chain fragment and nearby molecules.  In
particular, this field is affected by collective degrees of freedom of
non-absorbing units such as main chains and determines angular
distribution of the molecules in the stationary regime. It bears close
resemblance to the equilibrium distribution of the mean field theories
of POA. In Refs.~\cite{Ped:1997,Ped:1998,Hvil:2001} this distribution
has been assumed to be proportional to $\exp(-V(\uvc{n})/k_B T)$,
where $V(\uvc{n})$ is the mean-field potential that depends on the
averaged order parameter tensor.

So, we have the additional subsystem characterized by
$f_p(\uvc{n},t)$ attributed to the presence of long-living angular
correlations coming from anisotropic interactions between 
azochromophores 
and collective modes of polymeric environment.  For brevity, we
shall refer to the subsystem as a polymer system (matrix).  We shall
write the kinetic rate equations for $N_{\alpha}(\uvc{n},t)$ 
in the following form of master 
equations~\cite{Kis:epj:2001}:
\begin{align}
&    
\pdr{N_\alpha}=
\int
\Bigl[\,W(\alpha,\uvc{n}\,|\,\beta,\uvc{n}')
N_\beta(\uvc{n}',t)
\notag\\
&
-W(\beta,\uvc{n}'\,|\,\alpha,\uvc{n})\,
N_\alpha(\uvc{n},t)\,
\Bigr]\,\dd\uvc{n}'\notag\\
&
+\gamma_{\alpha}
\Bigl[\,N_{\alpha}(t)
\int\Gamma_{\alpha-p}(\uvc{n},\uvc{n}') f_p(\uvc{n}',t)\dd\uvc{n}'
\notag\\
&
- N_{\alpha}(\uvc{n},t)\,\Bigr]\,,\quad \alpha\ne\beta\,,
\label{eq:master}
\end{align}
where $\alpha\,, \beta\in\{tr,\,cis\}$.

The rate of 
\trans-\cis\  photoisomerization
stimulated by the incident UV light
enters the first bracketed term
on the right hand side of Eq.~\eqref{eq:master}.
For the electromagnetic wave linearly
polarized along the $x$-axis 
this rate can be written as follows~\cite{Dum:1992,Dum:1996}:
\begin{align}
&
W(cis,\uvc{n}\,|\,tr,\uvc{n}')=
\Gamma_{t-c}(\uvc{n},\uvc{n}')\,P_{tr}(\uvc{n}'),
  \label{eq:wc-t}\\
&
P_{tr}(\uvc{n})=(\hbar\omega_t)^{-1}\Phi_{tr\to cis}\sum_{i,j}
\sigma_{ij}^{(tr)}(\uvc{n})E_i E_j^{*}\notag\\
&
=q_t I (1+u\, n_x^2)\,,
\label{eq:ptr}
\end{align}
where 
$\bs{\sigma}^{(tr)}(\uvc{n})$ is the tensor of 
absorption cross section
for the \trans\  molecule oriented along $\uvc{n}$:
${\sigma}^{(tr)}_{ij}=\sigma_{\perp}^{(tr)}\delta_{ij}+
(\sigma_{||}^{(tr)}-\sigma_{\perp}^{(tr)})\,
n_i\,n_j$;
$
u\equiv
(\sigma_{||}^{(tr)}-\sigma_{\perp}^{(tr)})/\sigma_{\perp}^{(tr)}
$
is the absorption anisotropy parameter;
$\hbar\omega_t$ is the photon energy;
$\Phi_{tr\to cis}$ is the quantum yield of the process and 
$\Gamma_{t-c}(\uvc{n},\uvc{n}')$ describes the angular redistribution of
the molecules excited in the \cis\  state;
$I$ is the pumping intensity and
$q_t\equiv (\hbar\omega_t)^{-1}\Phi_{tr\to cis}\sigma_{\perp}^{(tr)}$.
 
Similarly, the rate of  \cis-\trans\ transition is
\begin{align}
  \label{eq:wt-c}
W(tr,\uvc{n}\,|\,cis,\uvc{n}')=
(\gamma_c+q_c I)\,\Gamma_{c-t}(\uvc{n},\uvc{n}')\,,
\end{align}
where 
$q_c\equiv (\hbar\omega_t)^{-1}\Phi_{cis\to tr}\sigma^{(cis)}$ and
$\gamma_c\equiv 1/\tau_c$, $\tau_c$ is the lifetime of \cis\ 
fragments and the anisotropic part of the absorption cross section is
disregarded,
$\sigma_{||}^{(cis)}=\sigma_{\perp}^{(cis)}\equiv\sigma^{(cis)}$.
There are angular redistribution probabilities 
that enter both Eqs.~\eqref{eq:wc-t} and~\eqref{eq:wt-c} and
meet the standard normalization condition for probability densities 
\begin{equation}
  \label{eq:norm}
  \int\Gamma_{\beta-\alpha}(\uvc{n},\uvc{n}')\,\dd\uvc{n}=1\, .
\end{equation}

From Eqs.~\eqref{eq:master},~\eqref{eq:wc-t} and~\eqref{eq:wt-c} it is
not difficult to deduce the equation for $n_{tr}(t)$:
\begin{equation}
  \label{eq:n-tr}
  \pdr{n_{tr}}=
(\gamma_c+q_c I)\,n_{cis}-\langle P_{tr}\rangle_{tr} n_{tr}\, ,
\end{equation}
where the angular brackets $\langle\ldots\rangle_\alpha$ 
stand for averaging over angles with the distribution function
$f_{\alpha}$ . Remarkably, this equation does not depend on the form of the
angular redistribution probabilities.

The last square bracketed term on the right hand side of
Eq.~\eqref{eq:master} describes 
the process that equilibrates 
the side chain absorbing molecules and the polymer system
in the absence of irradiation.
The angular redistribution probabilities
$\Gamma_{\alpha-p}(\uvc{n},\uvc{n}')$ meet the normalization
condition~\eqref{eq:norm}, so that  this thermal relaxation does not
change the total fractions $N_{tr}$ and $N_{cis}$. 
If there is no angular redistribution, 
then $\Gamma_{\alpha-p}(\uvc{n},\uvc{n}')=\delta(\uvc{n}-\uvc{n}')$
and both equilibrium angular distributions 
$f_{tr}^{(eq)}$ and $f_{cis}^{(eq)}$ are equal to $f_p$.

By contrast to the mean field theories,
where the mean-field potential is defined through
the self-consistency condition,
our approach is to
determine the distribution function $f_p(\uvc{n},t)$
from the kinetic equation written in the following form: 
\begin{align}
&
\pdr{f_{p}(\uvc{n},t)}= -\sum_{\alpha=\{tr,cis\}}
\gamma_p^{(\alpha)}\,n_{\alpha}(t)\,
\Bigl[\,f_p(\uvc{n},t) \notag\\
&
-\int\Gamma_{p-\alpha}(\uvc{n},\uvc{n}')
f_{\alpha}(\uvc{n}',t)\,\dd\uvc{n}'\,\Bigr]\,.
\label{eq:gen-p}
\end{align}
This equation combined with
Eqs.~\eqref{eq:master}-\eqref{eq:wt-c} 
can be used to formulate a number of phenomenological models of POA.
In particular, it can be shown~\cite{Kis:jpcm:2002} that
the results of 
Refs.~\cite{Ped:1997,Ped:1998,Hvil:2001,Puch:1998,Puch:1999}
can be recovered by choosing suitably defined angular redistribution
probabilities.

We can now describe our model. 
The first assumption is that all the
angular redistribution operators $\Gamma_{t-c}$
and $\Gamma_{c-t}$ take the following isotropic form:
\begin{align}
 \Gamma_{c-t}(\uvc{n},\uvc{n}')=
\Gamma_{t-c}(\uvc{n},\uvc{n}')=\frac{1}{4\pi}\equiv f_{iso}\,.
\label{eq:as-iso}
\end{align}
Since we have neglected anisotropy of \cis\ fragments 
in Eq.~\eqref{eq:wt-c},
it is reasonable to suppose that the equilibrium
distribution of \cis\ molecules is also isotropic,
$f_{cis}^{(eq)}=f_{iso}$.
So, we have
\begin{equation}
  \label{eq:as-gamma}
\gamma_{cis}=\gamma_p^{(cis)}=0\,.  
\end{equation}
We also assume that 
the equilibrium angular distribution of \trans\ fragments is determined
by the polymer system: $f_{tr}^{(eq)}=f_{p}$.
The latter will give the relation
\begin{equation}
  \label{eq:as-p}
  \Gamma_{\alpha-p}(\uvc{n},\uvc{n}')
=\Gamma_{p-\alpha}(\uvc{n},\uvc{n}')=\delta(\uvc{n}-\uvc{n}')\,.
\end{equation}
So, in this model equilibrium properties of
\cis\ and \trans\ isomers characterized by
two different equilibrium angular distributions:
$f_{iso}$ and $f_p$, respectively. It means that
the anisotropic field represented by $f_p$ does not influence
the angular distribution of non-mesogenic \cis\ fragments.
 
Eqs.~\eqref{eq:as-iso}--\eqref{eq:as-p} can now be used 
to obtain kinetic equations for the angular distribution functions of isomers 
$f_{\alpha}$ and of the polymer system $f_p$.
It is, however, more suitable 
to describe the temporal evolution of photoinduced
anisotropy in terms of 
the components of the order parameter tensor~\cite{Gennes:bk:1993}
\begin{equation}
  \label{eq:tens}
  S_{ij}(\uvc{n})=2^{-1}\,(3 n_i n_j -\delta_{ij})\, .
\end{equation}

Integrating the equations for the angular distribution 
functions multiplied by $S_{ij}(\uvc{n})$
over the angles will provide a set of equations
for the averaged order parameter components
$S_{ij}^{(\alpha)}\equiv\langle S_{ij}(\uvc{n})\rangle_\alpha$. 
The simplest result is for the order parameters of \cis\ molecules:
\begin{equation}
  \label{eq:ord-cis}
n_{cis}\pdr{S_{ij}^{(cis)}}=
-n_{tr}\langle P_{tr}\rangle_{tr}\ S_{ij}^{(cis)}\, .
\end{equation} 
From Eq.~\eqref{eq:ord-cis} initially isotropic (and equilibrium)
angular distribution of \cis\ fragments remains
unchanged in the course of irradiation.
In this case we have no effects due to the
ordering kinetics of \cis\ molecules.

The remaining part of the equations
for the order parameter components is given by
\begin{subequations}
\label{eq:gen-ord}
\begin{align}
&
n_{tr}\pdr{S_{ij}^{(tr)}}= -2/3\, q_t I u\,
n_{tr}\,G_{ij;\,xx}^{(tr)}-n_{cis}\notag\\ 
&
\times (\gamma_c+q_c I)\,S_{ij}^{(tr)} +
\gamma_{tr} n_{tr} (S_{ij}^{(p)}-S_{ij}^{(tr)})\, ,
\label{eq:gen-ord-tr}\\
&\pdr{S_{ij}^{(p)}}=-\gamma_{p} n_{tr}(S_{ij}^{(p)}-S_{ij}^{(tr)})\, ,
\label{eq:gen-ord-p}
\end{align}
\end{subequations}
where $\gamma_p\equiv\gamma_p^{(tr)}$ and 
$G_{ij;\,mn}^{(\alpha)}$ is the order parameter
correlation function (correlator) defined as follows:
\begin{equation}
  \label{eq:cor-fun}
  G_{ij;\,mn}^{(\alpha)}=
\langle S_{ij}(\uvc{n})S_{mn}(\uvc{n}) \rangle_\alpha-
S_{ij}^{(\alpha)}\,S_{mn}^{(\alpha)}\, .
\end{equation}
These functions characterize response of the side groups to the
pumping light.

Eqs.~\eqref{eq:gen-ord} will give the system for the components of the
order parameter tensor, if a closure can be found linking the
correlation functions and $S_{ij}^{(\alpha)}$.  The simplest closure
can be obtained by writing the products of the order parameter
components as a sum of spherical harmonics and neglecting the high
order harmonics with $j>2$, where $j$ is the angular momentum number.  
This procedure implies using the truncated expansion
of the distribution function $f_{tr}$ over the spherical harmonics
and leads to the parabolic approximation for the autocorrelators
of diagonal order parameter components~\cite{Kis:epj:2001}:
\begin{equation}
  \label{eq:decoup}
  G_{ii;\,ii}^{(tr)}\equiv G_{ii}\approx
1/5+2/7\,S_i-S_i^2\,,
\end{equation}
where $G_{ij}\equiv G_{ii;\,jj}^{(tr)}$ and 
$S_i\equiv\langle S_{ii}\rangle_{tr}$.

It is known~\cite{Gennes:bk:1993} that
the values of $S_i$ are varied from $-0.5$ to $1$ and
the approximate expressions~\eqref{eq:decoup}
can be justified only if $S_i$ is sufficiently small
for the contributions of the high order harmonics
to be negligible. Otherwise, Eq.~\eqref{eq:decoup}
predicts negative values for 
non-negatively defined autocorrelators $G_{ii;\,ii}^{(tr)}$ 
when $S_i$ is in the vicinity of the end points $-0.5$ and $1$.
This would lead to physically absurd results and 
the parabolic approximation severely breaks down. 

In order to restore the correct behavior, 
we need either to modify the approximation~\eqref{eq:decoup}
or to deal with a system of coupled equations for 
high order harmonics. The latter involves
truncating the expansion for the distribution function $f_{tr}$ 
at sufficiently large $j$.
But using this approach to treat the case of order parameters
$S_i$  close to the end points 
would require large scale computations complicated by 
the convergence and stability problems.

In this paper
we shall use an alternative procedure by
assuming that the angular distribution of mesogenic
groups in azopolymers can be taken in the form
of distribution functions used in the variational mean field theories
of liquid crystals~\cite{Gennes:bk:1993,Luben:bk:1995}
\begin{align}
&
  f=N^{-1} \exp\left(\sum_{i,j}c_{ij} S_{ij}(\uvc{n})\right),
  \label{eq:MSp}
\\
&
N=4\pi\int_0^1\exp[(c_1-(c_2+c_3)/2)(3\tau^2-1)/2]
\notag\\
&
\times
I_0\left(3(c_3-c_2)(1-\tau^2)/4\right)\dd\tau,
\label{eq:MSp-norm}
\end{align}
where $N$ is the normalization coefficient, 
$c_i$ are the eigenvalues of the tensor $c_{ij}$ and 
$I_0(x)$ is the modified Bessel
function of the zero order~\cite{Abr}.

We can now consider the simplest modification
of the parabolic approximation obtained
by rescaling the
order parameter components: $\langle S_{ii}\rangle_{tr}\to
\lambda\langle S_{ii}\rangle_{tr}$  
and use the mean field parameterization~\eqref{eq:MSp}
to estimate its accuracy.
The coefficient $\lambda$ is
computed from the condition that there are
no fluctuations provided the molecules are perfectly aligned 
along the coordinate unit vector $\uvc{e}_i$:
$G_{ii;\,ii}^{(tr)}=0$ at
$\langle S_{ii}\rangle_{tr}=1$. It  gives the value
of $\lambda$ equal to $(1+0.6\sqrt{30})/7$.
The results of
numerical analysis reported in~\cite{Kis:jpcm:2002} 
shows that this heuristic procedure 
gives reasonably accurate approximation for correlators of the
mean field distributions~\eqref{eq:MSp}.

By using the modified parabolic approximation for the correlators,
we can now write down  
the resulting system for the diagonal components of
the order parameter tensor in the final form:
\begin{align}
&n_{tr}\pdr{S}= -2u/3\, q_t I (5/7+2\lambda/7\,S-\lambda^2 S^2) n_{tr}\notag\\
&-(\gamma_c+q_c I)\, n_{cis} S +\gamma_{tr} n_{tr} (S_p-S),
\label{eq:ord-tr1}
\end{align}
\begin{align}
&n_{tr}\pdr{\Delta S}= 2u/3\, q_t I\lambda (2/7+\lambda S) n_{tr}\Delta S 
-n_{cis}\notag\\
&\times (\gamma_c+q_c I)\Delta S +
\gamma_{tr} n_{tr} (\Delta S_p-\Delta S),
\label{eq:ord-tr2}
\end{align}
\begin{align}
&\pdr{S_p}=-\gamma_{p} n_{tr} (S_p-S)\, ,
\label{eq:ord-p1}
\end{align}
\begin{align}
&\pdr{\Delta S_p}=-\gamma_{p} n_{tr} (\Delta S_p-\Delta S)\, ,
\label{eq:ord-p2}
\end{align}
where $S\equiv \langle S_{xx}\rangle_{tr} $,
$\Delta S\equiv \langle S_{yy}-S_{zz}\rangle_{tr}$,
$S_p\equiv \langle S_{xx}\rangle_{p} $ and
$\Delta S_p\equiv \langle S_{yy}-S_{zz}\rangle_{p}$.

In  polymer P1 the part of the
correlation functions responsible for
out-of-plane reorientation is appeared to be
suppressed by polymeric environment.
As in Ref.~\cite{Kis:epj:2001} we shall account for the presence
of these constraints by assuming that
$G_{ij;\,xx}^{(tr)}\approx -G_{ij;\,yy}^{(tr)}$.
The result is that the equations for the diagonal 
order parameter components will be in the form of 
Eqs.~\eqref{eq:ord-tr1}-~\eqref{eq:ord-p2}, where
we need to change the sign of the first term on the right
hand sides of Eqs.~\eqref{eq:ord-tr1}-~\eqref{eq:ord-tr2}
and to interchange $S_{xx}$ and $S_{yy}$.

\section{Numerical results}
\label{sec:num-res}

In this section we employ our model to
interpret the experimental data of the UV absorption measurements for
different irradiation doses.
The principal absorption coefficients $D_i$ can be related to the
concentrations and the order parameters as follows
\begin{align}
  &
  D_i\propto \langle\sigma_{ii}^{(tr)}\rangle_{tr}\,n_{tr}
+\sigma^{(cis)}\, n_{cis}\notag\\
&
\propto (1+u^{(a)}(2\,S_i+1)/3) n_{tr}+q_{ct} n_{cis}\,,
\label{eq:abs_coef}
\end{align}
where $S_i\equiv \langle S_{ii}\rangle_{tr}$;
$u^{(a)}$ is the absorption anisotropy parameter and 
$q_{ct}$ is the ratio of $\sigma^{(cis)}$ and
$\sigma_{\perp}^{(tr)}$ at the wavelength of probing light.

Eq.~\eqref{eq:abs_coef} implies that
the method of total absorption based on Eq.~\eqref{eq:D_tot}
is applicable when the fraction of \cis\ fragments is negligible,
$n_{cis}\approx 0$. In this case from Eq.~\eqref{eq:abs_coef} 
the absorption
order parameters~\eqref{eq:S_abs} are proportional to $S_i$:
$S_i^{(a)}=u^{(a)}/(3+u^{(a)}) S_i$.

The theoretical curves are computed by solving the kinetic equations
of the model discussed in the previous section.  Initial values of the order
parameters $S(0)$ and $\Delta S(0)$ are taken from the experimental
data measured at $\lambda_{ex}=488$~nm.  Since the system is initially
at the equilibrium state, the remaining part of the initial conditions
is: $S_p(0)=S(0)$, $\Delta S_p(0)=\Delta S(0)$, $n_{tr}(0)=1$ and
$n_{cis}(0)=0$.

Anisotropy of non-irradiated films is uniaxial for both polymers:
$S_z^{(a)}(0)=-0.18<S_x^{(a)}(0)=S_y^{(a)}(0)=0.09$ (polymer P1)
and $S_z^{(a)}(0)=0.07>S_x^{(a)}(0)=S_y^{(a)}(0)=-0.035$
(polymer P2). From the other hand, reaching the photosteady state
orientational structure of the polymers is characterized by
different uniaxial anisotropies:
$S_x^{(st)}=S_z^{(st)} < S_y^{(st)}$ (polymer P1)
and $S_x^{(st)} < S_y^{(st)} = S_z^{(st)}$ (polymer P2).
So, as is shown in Figs.~\ref{fig:P1-365}--~\ref{fig:P2-488},
the transient anisotropic structures are inevitably biaxial.
These biaxial effects are related to the difference
between the initial anisotropy of polymer films and the 
anisotropy of the photo-saturated state.

Numerical calculations in the presence of irradiation were followed by
computing the stationary values of $S$ and $\Delta S$ to which the
order parameters decay after switching off the irradiation at time
$t_0$. For the polymer P2
the lifetime of the \cis\ fragments was found to be much longer than
the periods examined and we can safely neglect $\gamma_c$.
If $\gamma_c=0$ the kinetic equations in the absence of irradiation
can be easily solved to yield the stationary value of
$S_i$ and $S_i^{(p)}$:
$\bigl(\gamma_p n_{tr}(t_0) S_i(t_0) + 
\gamma_{tr} S_i^{(p)}(t_0)\bigr)/\gamma$,
where $\gamma\equiv\gamma_p n_{tr}(t_0) + \gamma_{tr}$.
There is no further relaxation after reaching this stationary state
and its anisotropy is long term stable.
The lifetime of \cis\ forms in polymer P1 is $\tau_c=1.2$~s
and the kinetic equations need to be solved numerically.
This, however, does not affect the conclusion about the long term
stability of POA.

According to Ref.~\cite{Puch:1998},
the relaxation time characterizing
decay of $D_{i}(t)$ to its stationary value after switching off the
irradiation in polymer P1 can be estimated at about $\tau_t=15$~min. 
The experimental estimate for polymer P2 is $\tau_t=30$~min.  
The theoretical value of this relaxation
time, deduced from solution of the kinetic equations in the absence of
irradiation, is $\tau_t\approx 1/(\gamma_p + \gamma_{tr})$.  
So, in the simplest
case, we can assume both relaxation times, $\tau_p$
($\gamma_p=1/\tau_p$) and $\tau_{tr}$ ($\gamma_{tr}=1/\tau_{tr}$), to
be equal 30~min (P1) and 60~min (P2).
Table~\ref{tab:param-pol} shows the estimates for
the absorption cross section of \cis\ molecules 
$\sigma^{(cis)}$ and 
the average absorption cross section of \trans\ fragments,
$\sigma^{(tr)}=(\sigma_{||}^{(tr)}+2\sigma_{\perp}^{(tr)})/3$,
obtained from the UV spectra of the polymers dissolved in toluene.

For these polymers the absorption anisotropy parameters and the quantum
efficiencies are unknown and need to be fitted.  We used the value of
$S_{st}$ as an adjustable parameter, so that the anisotropy parameters
$u$ and $u^{(a)}$ can be derived from Eq.~\eqref{eq:S_st} and from the
experimental value of the absorption order parameter $S_{st}^{(a)}$
measured at $\lambda_{ex}=488$~nm in the photosteady state.  

The numerical results presented in Figs.~\ref{fig:P1-365}-\ref{fig:P2-488}
are computed at $u^{(a)}=13.0$ and $q_{ct}=2.9$ (polymer P1);
$u^{(a)}=11.0$ and $q_{ct}=2.15$ (polymer P2).
The quantum efficiencies are listed in Table~\ref{tab:param-pol} and 
are of the same order of magnitude
as the experimental values for other azobenzene
compounds~\cite{Mita:1989}.
  
In order to characterize the regime of POA,
we can use the fraction of \cis\ fragments in the
photo-stationary state.
From Eq.~\eqref{eq:n-tr} this fraction is given by
\begin{equation}
n_{cis}^{(st)}=\frac{3+u(1+2 S_{st})}{3(r+1)+u(1+2 S_{st})}\,,
  \label{eq:nc_st}
\end{equation}
where $r\equiv (\gamma_c+q_c I)/(q_t I)$,
$S_{st}\equiv S_x^{(st)}$ and
the corresponding value of the order parameter is a solution
of the following equation  
\begin{align}
&
2 u\,(1/5\,+2\lambda/7\, S_{st}-\lambda^2 S_{st}^{\,2})\notag\\
&
=-S_{st}(3+u(1+2 S_{st}))\,,
  \label{eq:S_st}
\end{align}
deduced by using Eqs.~\eqref{eq:ord-tr1}
and~\eqref{eq:ord-p1}. 

At small values of $r$ Eq.~\eqref{eq:nc_st} will yield the
fraction $n_{cis}^{(st)}$ that is close to the unity and we have the
kinetics of POA in the regime of photoselection.  In the opposite case
of sufficiently large values of $r$ the photosteady fraction of \cis\
molecules will be very small that is typical of the photo-reorientation
mechanism.  

When, as in polymer P1, \cis\  isomers are short-living,
so that the rate of thermal relaxation $\gamma_c$ 
is relatively large, the values of
the parameter $r$ listed in Table~\ref{tab:param-pol}
are large at both wavelengths of the exciting light. 
In this case the fractions of \cis\ isomers 
in the photo-saturated state computed from
Eqs.~\eqref{eq:nc_st}-~\eqref{eq:S_st} are negligibly small:
$n_{cis}^{(st)}\approx 0.003$ at $\lambda_{ex}=365$~nm and
$n_{cis}^{(st)}\approx 0.002$ at $\lambda_{ex}=488$~nm.
So, we have the kinetics of POA in the regime of
photo-reorientation. It is illustrated in 
Figs.~\ref{fig:P1-365}(b) and~\ref{fig:P1-488}(b),
where the measured and the calculated curves 
are in good agreement and correspond to
the photo-reorientation mechanism.

In polymer P2 \cis\  isomers are long-living with
negligibly small relaxation rate $\gamma_c$. 
The parameter $r$ is now the
ratio of the rates $q_c I$ and $q_t I$
that characterize $cis \to trans$ and $trans \to cis$
stimulated transitions, correspondingly.
This ratio is determined by the quantum efficiencies,
$\Phi_{tr\to cis}$ and $\Phi_{cis\to tr}$,
and the absorption cross sections,
$\sigma_{\perp}^{(tr)}$ and $\sigma_{cis}$.
In this case we can have both mechanisms depending on
the wavelength of excitation.

Fig.~\ref{fig:P2-488} 
presents the case in which the wavelength of
light is far from the absorption maximum, 
$\lambda_{ex}=488$~nm.
It seen that, similar to polymer P1, 
the kinetics of the absorption coefficients,
presented in Fig.~\ref{fig:P2-488}(b), 
is typical for photo-reorientation mechanism.
According to Table~\ref{tab:param-pol}, the value of $r$
is $ 69.1$ and we can use Eqs.~\eqref{eq:nc_st}-~\eqref{eq:S_st}
to estimate the fraction $n_{cis}^{(st)}$ at about $0.02$.
It means that in this case the $cis\to trans$ transitions stimulated
by the exciting light will efficiently deplete the \cis\ state
and the absorption coefficients are controlled by the terms
proportional to the order parameter of \trans\ molecules
(see Eq.~\eqref{eq:abs_coef}).  
 
By contrast, Fig.~\ref{fig:P2-365}(b) shows that
the experimental dependencies for both $D_x$ and $D_y$
are decreasing functions of the irradiation time
at the wavelength, $\lambda_{ex}=365$~nm, near the maximum of absorption 
band. 
It indicates that the kinetics of 
POA is governed by the mechanism of photoselection.
In this case we have $r=0.03$ and $n_{cis}^{(st)}=0.98$.
The method described in Sec.~\ref{subsec:abs-method}
is now inapplicable, so the out-of-plane absorption
coefficient $D_z$ and the order parameters $S_i^{(a)}$
can only be estimated theoretically.
These curves calculated from our model
are shown in Figs.~\ref{fig:P2-365}(b)-(c). 

\section{Discussion and Conclusions}
\label{sec:discussion}

In Sec.~\ref{sec:num-res} we have introduced 
the parameter $r$ that can be written as the ratio
of the characteristic time of \trans-\cis\ isomerization,
$\tau_{ex}=1/(q_t I)$, to the time characterizing
decay of the \cis\ state in the presence of irradiation,
$\tilde\tau_c=1/(\gamma_c+q_c I)$.
We have also shown that depending on the value of this parameter
the kinetics of POA is governed by either photoselection
or photo-reorientation mechanisms. 

For large values of $r$, $r\gg 1$, 
we have $\tilde\tau_c\ll\tau_{ex}$, so that
the lifetime of \cis\ isomers under irradiation $\tilde\tau_c$
is short and the isomers are short-living.  
In this case the photo-reorientation
mechanism is found to dominate the kinetics of POA.
The opposite case of small values of $r$ 
characterizes the regime of photoselection,
where $\tilde\tau_c\gg\tau_{ex}$ and 
\cis\ isomers are long-living in the presence of irradiation.

It should be stressed that the mechanism of photoselection
cannot occur in polymers with high rate of thermal
\cis-\trans\ isomerization $\gamma_c$, where
the time of \trans-\cis\ isomerization
$\tau_{ex}$ is longer than
the lifetime of \cis\ isomers $\tau_c=1/\gamma_c > \tilde\tau_c$. 
In our study this case is represented
by polymer P1 in which the mechanism of photo-reorientation
dominates at both wavelengths of excitation.

In polymer P2 $\gamma_c$ is very small and the relaxation time
$\tilde\tau_c$ is determined by the rate of
\cis-\trans\ photoisomerization $q_c I$.
When the wavelength of the exciting light is in 
n$\pi^{*}$ absorption band of
azochromophores, this rate is higher than
the rate of \trans-\cis\ isomerization,
$q_t I$.  So, at $\lambda_{ex}=488$~nm 
the photo-reorientation mechanism 
dominates in both polymers.

By contrast to n$\pi^{*}$ absorption band, $\pi\pi^{*}$ absorption
bands of isomers are well separated.  The wavelength
$\lambda_{ex}=365$~nm lies within $\pi\pi^{*}$ absorption band of
\trans\ chromophores, so that the rate $q_c I$ is low as compared to
$q_t I$.  For polymer P2 this means that
there is nothing to prevent the \cis\ 
state from being populated under the action of UV light and the
kinetics of POA is now governed by the mechanism of photoselection.

The photochemical properties of azopolymers are determined in large
part by the molecular structure of azochromophores, which incorporates
the azobenzene core and substitutes.  For instance, sufficiently
polarized azochromophores (push-pull chromophores) usually have
short-living \cis\ isomers.  On the other hand, the chromophores
containing long alkyl substitutes are characterized by long-living
\cis\ form.  Our results show that the polymers under consideration
comply with these rules.

In addition, the molecular structure of azopolymers determines
self-organization of azochromophores, which in turn affects 3D
orientational ordering. By self-organization is meant a
number of complex processes related to the spontaneous alignment and
aggregation of anisotropic azobenzene chromophores, processes of
collective orientation, self-assembling at the interfaces etc.  The
initial anisotropy of non-irradiated polymer films is defined by
these processes.  We found that the azochromophores in P1 films prefer
in-plane orientation, whereas the preferential alignment of the
fragments in P2 films is homeotropic.  
The factors responsible for this
difference have not been studied in any detail yet.

The intrinsic self-organization of azochromophores also 
plays a part in the ordering stimulated by actinic light~\cite{Zakr:2001}. 
The excitation of
azochromophores stimulates these  processes of self-organization
that are slowed down in the glassy state. The contribution of
self-organization processes may explain different anisotropies
observed in polymers P1 and P2 on reaching the state of photo-saturation. 
The positive in-plane order, observed
in polymer P1, is determined by the strong
preference of azochromophores to in-plane alignment. 

The negative
in-plane order induced in polymer P2 by the light with
$\lambda_{ex}=488$~nm  may be
explained by assuming the dominating role of the photo-reorientation
mechanism. 
Generally, we know from experience that
the negative uniaxial anisotropy with the optical axis
parallel to the polarization vector of the exciting light
can be easily induced in polymers showing preference to out-of-plane
alignment of azochromophores.

The photosaturated state of polymer P2 films 
irradiated at $\lambda_{ex}=365$~nm is found to 
be optically isotropic with $n_x=n_y=n_z$. 
As we have already noted in Sec.~\ref{subsubsec:p2-365}, 
extremely low concentration of \trans\ chromophores will 
render the film effectively isotropic. This is the case even
if the angular distribution of \trans\ isomers 
retains an amount of anisotropy caused by the angular selective character
of the burning process. 
From the other hand, non-mesogenic \cis\ isomers at 
high concentrations could result in disordering effects 
up to suppressing LC properties of azopolymers~\cite{Kanaz:1997}.
From our optical measurements we cannot 
unambiguously conclude on significance of these effects for polymer P2. 

Thus, in the regime of photo-reorientation
the orientational structures observed in the photo-saturation
state are uniaxial with optical axes
determined by the polarization of the
light and by the favored orientation of the azochromophores.  In
the case of photoselection these structures are optically isotropic.
The anisotropic structures induced at early stages of irradiation are
biaxial.  These transient structures are formed in passing from the
initial uniaxial structure to the state of photo-saturation.

Self-organization of azochromophores under irradiation implies their
collective orientation and liquid crystallinity. 
The other collective mode is related to the orientation of
non-photosensitive fragments (matrix) that is influenced by the orientation of
azochromophores.  According to Refs.~\cite{Puch:1998,Puch:1999}, the
latter factor seems to be of crucial importance in stabilizing the
photoinduced order.

In our experiments the photoinduced anisotropy
has been seen to be stable over at least several weeks. 
The anisotropy can be erased and rewritten by the light. 
This long term stability of the induced order implies effective
orientation of polymer matrix in both photo-reorientation and
photoselection regimes of POA.

In order to interpret
the experimental results on the kinetics of the photoinduced
absorption dichroism we employed the theoretical
model formulated by using the phenomenological approach of 
Refs.~\cite{Kis:epj:2001,Kis:jpcm:2002}.
This approach describes the kinetics of POA in terms of
one-particle angular distribution functions and angular
redistribution probabilities. 
The probabilities enter the photoisomerization
rates and define coupling between the azochromophores and 
the anisotropic field represented by the distribution function of
the polymer matrix $f_p$. This anisotropic field
reflects the presence of long-living angular correlations
and stabilizes the photoinduced anisotropy. 

The key assumption of the model
is that the \cis\ fragments are isotropic and do not affect the
ordering kinetics directly. Certainly, this is the simplest case to
start from before studying more complicated models.
So, we studied the predictions of this simple model for both polymers
to test its applicability. To this end we have
estimated a number of photochemical parameters that enter the model
from the experimental data. Only the absorption anisotropy parameters
and the quantum yields need to be adjusted.
The comparison between the numerical results and the experimental data
shows that the model correctly captures the basic features of POA in
azopolymers.   

We have additionally calculated the out-of-plane absorbance $D_z$ and
the absorption order parameters $S_i^{(a)}$ in the photoselection
regime where the experimental method of estimation is inapplicable.
The model has also been applied to estimate concentrations of isomers,
quantum yields of photoisomerization and anisotropy of molecular
absorption.  It provides a criterion for the occurrence of different
mechanisms and describes the biaxiality effects.
 
So, we have demonstrated that the phenomenological approach
of~\cite{Kis:epj:2001,Kis:jpcm:2002} can be used as a useful tool for
studying photoinduced ordering processes in azopolymers.  But it
should be noted that theoretical approaches of this sort, by
definition, do not involve explicit considerations of microscopic
details of azopolymer physics.  A more comprehensive study is
required to relate the effective parameters of our model and physical
parameters characterizing interactions between molecular units of
polymers.  For example, the initial anisotropy of polymer films is
taken into account through the initial conditions for the kinetic
equations and the theory then properly describes the biaxiality
effects.  But the process of structure formation in non-irradiated
polymer films is well beyond the scope of the model and such
description cannot serve as an explanation of biaxiality.  Similarly,
the suppressed out-of-plane reorientation of azochromophores in
polymer P1 is treated as a constraint imposed by the polymeric
environment and is taken into consideration by modifying the order
parameter correlation functions.

Our results show that neglecting the influence of \cis\ isomers on the
orientational order of the anisotropic \trans\ isomers does not lead
to considerable discrepancies between the theoretical results and the
experimental data. In general, the presence of non-mesogenic \cis\
isomers could give rise to deterioration of the photoinduced order of
\trans\ fragments. In the regime of photoselection
this effect will eventually produce nearly isotropic 
angular distribution of \trans\ isomers.
On the other hand, 
our model predicts that in polymer P2 this distribution 
is always anisotropic with non-vanishing
order parameters of \trans\ fragments.
(In this case the order parameters are not proportional
to $S_i^{(a)}$ depicted in Fig.~\ref{fig:P2-365}(c)
that go to zero due to depletion of the \trans\ state.)
Though the model is successful in describing our experiments, 
it can be modified to study the case in which 
the disordering effect caused by \cis\
isomers cannot be disregarded. 

\begin{acknowledgments}
This study was partially supported by CRDF (grant UP1-2121B). We
also thank Dr. T. Sergan  from Kent State University
for assistance with processing the data of the null ellipsometry
measurements and fruitful discussion.
 
\end{acknowledgments}


\begin{thebibliography}{46}
\expandafter\ifx\csname natexlab\endcsname\relax\def\natexlab#1{#1}\fi
\expandafter\ifx\csname bibnamefont\endcsname\relax
  \def\bibnamefont#1{#1}\fi
\expandafter\ifx\csname bibfnamefont\endcsname\relax
  \def\bibfnamefont#1{#1}\fi
\expandafter\ifx\csname citenamefont\endcsname\relax
  \def\citenamefont#1{#1}\fi
\expandafter\ifx\csname url\endcsname\relax
  \def\url#1{\texttt{#1}}\fi
\expandafter\ifx\csname urlprefix\endcsname\relax\def\urlprefix{URL }\fi
\providecommand{\bibinfo}[2]{#2}
\providecommand{\eprint}[2][]{\url{#2}}

\bibitem[{\citenamefont{Weigert}(1919)}]{Wieg:1919}
\bibinfo{author}{\bibfnamefont{F.}~\bibnamefont{Weigert}},
  \bibinfo{journal}{Verh. d. phys. Ges.} \textbf{\bibinfo{volume}{21}},
  \bibinfo{pages}{485} (\bibinfo{year}{1919}).

\bibitem[{\citenamefont{Neporent and Stolbova}(1963)}]{Nep:1963}
\bibinfo{author}{\bibfnamefont{B.~S.} \bibnamefont{Neporent}} \bibnamefont{and}
  \bibinfo{author}{\bibfnamefont{O.~V.} \bibnamefont{Stolbova}},
  \bibinfo{journal}{Sov. Opt. Spectrosc.} \textbf{\bibinfo{volume}{14}},
  \bibinfo{pages}{624} (\bibinfo{year}{1963}).

\bibitem[{\citenamefont{Todorov et~al.}(1984)\citenamefont{Todorov, Tomova, and
  Nikolova}}]{Tod:1984}
\bibinfo{author}{\bibfnamefont{T.}~\bibnamefont{Todorov}},
  \bibinfo{author}{\bibfnamefont{N.}~\bibnamefont{Tomova}}, \bibnamefont{and}
  \bibinfo{author}{\bibfnamefont{L.}~\bibnamefont{Nikolova}},
  \bibinfo{journal}{Appl. Optics} \textbf{\bibinfo{volume}{23}},
  \bibinfo{pages}{4309} (\bibinfo{year}{1984}).

\bibitem[{\citenamefont{Eich et~al.}(1987)\citenamefont{Eich, Wendorff, Reck,
  and Ringsdorf}}]{Eich:1987}
\bibinfo{author}{\bibfnamefont{M.}~\bibnamefont{Eich}},
  \bibinfo{author}{\bibfnamefont{J.~H.} \bibnamefont{Wendorff}},
  \bibinfo{author}{\bibfnamefont{B.}~\bibnamefont{Reck}}, \bibnamefont{and}
  \bibinfo{author}{\bibfnamefont{H.}~\bibnamefont{Ringsdorf}},
  \bibinfo{journal}{Makromol. Chem., Rap. Commun.}
  \textbf{\bibinfo{volume}{8}}, \bibinfo{pages}{59} (\bibinfo{year}{1987}).

\bibitem[{\citenamefont{Natansohn et~al.}(1992)\citenamefont{Natansohn, Xie,
  and Rochon}}]{Nat:1992}
\bibinfo{author}{\bibfnamefont{A.}~\bibnamefont{Natansohn}},
  \bibinfo{author}{\bibfnamefont{S.}~\bibnamefont{Xie}}, \bibnamefont{and}
  \bibinfo{author}{\bibfnamefont{P.}~\bibnamefont{Rochon}},
  \bibinfo{journal}{Macromolecules} \textbf{\bibinfo{volume}{25}},
  \bibinfo{pages}{5531} (\bibinfo{year}{1992}).

\bibitem[{\citenamefont{Holme et~al.}(1996)\citenamefont{Holme, Ramanujam, and
  Hvilsted}}]{Holme:1996}
\bibinfo{author}{\bibfnamefont{N.~C.~R.} \bibnamefont{Holme}},
  \bibinfo{author}{\bibfnamefont{P.~S.} \bibnamefont{Ramanujam}},
  \bibnamefont{and} \bibinfo{author}{\bibfnamefont{S.}~\bibnamefont{Hvilsted}},
  \bibinfo{journal}{Applied Optics} \textbf{\bibinfo{volume}{35}},
  \bibinfo{pages}{4622} (\bibinfo{year}{1996}).

\bibitem[{\citenamefont{Petry et~al.}(1993)\citenamefont{Petry, Kummer,
  Anneser, Feiner, and Br\"{a}uchle}}]{Petry:1993}
\bibinfo{author}{\bibfnamefont{A.}~\bibnamefont{Petry}},
  \bibinfo{author}{\bibfnamefont{S.}~\bibnamefont{Kummer}},
  \bibinfo{author}{\bibfnamefont{H.}~\bibnamefont{Anneser}},
  \bibinfo{author}{\bibfnamefont{F.}~\bibnamefont{Feiner}}, \bibnamefont{and}
  \bibinfo{author}{\bibfnamefont{C.}~\bibnamefont{Br\"{a}uchle}},
  \bibinfo{journal}{Ber. Bunsenges. Phys. Chem.} \textbf{\bibinfo{volume}{97}},
  \bibinfo{pages}{1281} (\bibinfo{year}{1993}).

\bibitem[{\citenamefont{Wiesner et~al.}(1992)\citenamefont{Wiesner, Reynolds,
  Boeffel, and Spiess}}]{Wies:1992}
\bibinfo{author}{\bibfnamefont{U.}~\bibnamefont{Wiesner}},
  \bibinfo{author}{\bibfnamefont{N.}~\bibnamefont{Reynolds}},
  \bibinfo{author}{\bibfnamefont{C.}~\bibnamefont{Boeffel}}, \bibnamefont{and}
  \bibinfo{author}{\bibfnamefont{H.~W.} \bibnamefont{Spiess}},
  \bibinfo{journal}{Liq. Cryst.} \textbf{\bibinfo{volume}{11}},
  \bibinfo{pages}{251} (\bibinfo{year}{1992}).

\bibitem[{\citenamefont{Blinov et~al.}(1998)\citenamefont{Blinov, Kozlovsky,
  Ozaki, Skarp, and Yoshino}}]{Blin:1998}
\bibinfo{author}{\bibfnamefont{L.}~\bibnamefont{Blinov}},
  \bibinfo{author}{\bibfnamefont{M.}~\bibnamefont{Kozlovsky}},
  \bibinfo{author}{\bibfnamefont{M.}~\bibnamefont{Ozaki}},
  \bibinfo{author}{\bibfnamefont{K.}~\bibnamefont{Skarp}}, \bibnamefont{and}
  \bibinfo{author}{\bibfnamefont{K.}~\bibnamefont{Yoshino}},
  \bibinfo{journal}{J. Appl. Phys.} \textbf{\bibinfo{volume}{84}},
  \bibinfo{pages}{3860} (\bibinfo{year}{1998}).

\bibitem[{\citenamefont{Yaroshchuk
  et~al.}(2001{\natexlab{a}})\citenamefont{Yaroshchuk, Sergan, Lindau, Lee,
  Kelly, and Chien}}]{Yar:2001}
\bibinfo{author}{\bibfnamefont{O.}~\bibnamefont{Yaroshchuk}},
  \bibinfo{author}{\bibfnamefont{T.}~\bibnamefont{Sergan}},
  \bibinfo{author}{\bibfnamefont{J.}~\bibnamefont{Lindau}},
  \bibinfo{author}{\bibfnamefont{S.~N.} \bibnamefont{Lee}},
  \bibinfo{author}{\bibfnamefont{J.}~\bibnamefont{Kelly}}, \bibnamefont{and}
  \bibinfo{author}{\bibfnamefont{L.-C.} \bibnamefont{Chien}},
  \bibinfo{journal}{J. Chem. Phys.} \textbf{\bibinfo{volume}{114}},
  \bibinfo{pages}{5330} (\bibinfo{year}{2001}{\natexlab{a}}).

\bibitem[{\citenamefont{Blinov and Chigrinov}(1994)}]{Blin:b:1994}
\bibinfo{author}{\bibfnamefont{L.~M.} \bibnamefont{Blinov}} \bibnamefont{and}
  \bibinfo{author}{\bibfnamefont{V.~G.} \bibnamefont{Chigrinov}},
  \emph{\bibinfo{title}{Electrooptic effects in liquid crystal materials}}
  (\bibinfo{publisher}{Springer-Verlag}, \bibinfo{address}{Berlin},
  \bibinfo{year}{1994}).

\bibitem[{\citenamefont{Dumont and Sekkat}(1992)}]{Dum:1992}
\bibinfo{author}{\bibfnamefont{M.}~\bibnamefont{Dumont}} \bibnamefont{and}
  \bibinfo{author}{\bibfnamefont{Z.}~\bibnamefont{Sekkat}},
  \bibinfo{journal}{Proc. SPIE} \textbf{\bibinfo{volume}{1774}},
  \bibinfo{pages}{188} (\bibinfo{year}{1992}).

\bibitem[{\citenamefont{Dumont et~al.}(1994)\citenamefont{Dumont, Hosotte,
  Froc, and Sekkat}}]{Dum:spie:1994}
\bibinfo{author}{\bibfnamefont{M.}~\bibnamefont{Dumont}},
  \bibinfo{author}{\bibfnamefont{S.}~\bibnamefont{Hosotte}},
  \bibinfo{author}{\bibfnamefont{G.}~\bibnamefont{Froc}}, \bibnamefont{and}
  \bibinfo{author}{\bibfnamefont{Z.}~\bibnamefont{Sekkat}},
  \bibinfo{journal}{Proc. SPIE} \textbf{\bibinfo{volume}{2042}},
  \bibinfo{pages}{2} (\bibinfo{year}{1994}).

\bibitem[{\citenamefont{Puchkovs'ka et~al.}(1998)\citenamefont{Puchkovs'ka,
  Reshetnyak, Tereshchenko, Yaroshchuk, and Lindau}}]{Puch:1998}
\bibinfo{author}{\bibfnamefont{G.~A.} \bibnamefont{Puchkovs'ka}},
  \bibinfo{author}{\bibfnamefont{V.~Y.} \bibnamefont{Reshetnyak}},
  \bibinfo{author}{\bibfnamefont{A.~G.} \bibnamefont{Tereshchenko}},
  \bibinfo{author}{\bibfnamefont{O.~V.} \bibnamefont{Yaroshchuk}},
  \bibnamefont{and} \bibinfo{author}{\bibfnamefont{J.}~\bibnamefont{Lindau}},
  \bibinfo{journal}{Mol. Cryst. Liq. Cryst.} \textbf{\bibinfo{volume}{321}},
  \bibinfo{pages}{31} (\bibinfo{year}{1998}).

\bibitem[{\citenamefont{Zakrevskyy et~al.}(2001)\citenamefont{Zakrevskyy,
  Yaroshchuk, Stumpe, Lindau, Sergan, and Kelly}}]{Zakr:2001}
\bibinfo{author}{\bibfnamefont{Y.}~\bibnamefont{Zakrevskyy}},
  \bibinfo{author}{\bibfnamefont{O.}~\bibnamefont{Yaroshchuk}},
  \bibinfo{author}{\bibfnamefont{J.}~\bibnamefont{Stumpe}},
  \bibinfo{author}{\bibfnamefont{J.}~\bibnamefont{Lindau}},
  \bibinfo{author}{\bibfnamefont{T.}~\bibnamefont{Sergan}}, \bibnamefont{and}
  \bibinfo{author}{\bibfnamefont{J.}~\bibnamefont{Kelly}},
  \bibinfo{journal}{Mol. Cryst. Liq. Cryst.} \textbf{\bibinfo{volume}{365}},
  \bibinfo{pages}{415} (\bibinfo{year}{2001}).

\bibitem[{\citenamefont{Buffeteau and P\'{e}zolet}(1998)}]{Buff:1998}
\bibinfo{author}{\bibfnamefont{T.}~\bibnamefont{Buffeteau}} \bibnamefont{and}
  \bibinfo{author}{\bibfnamefont{M.}~\bibnamefont{P\'{e}zolet}},
  \bibinfo{journal}{Macromolecules} \textbf{\bibinfo{volume}{31}},
  \bibinfo{pages}{2631} (\bibinfo{year}{1998}).

\bibitem[{\citenamefont{Kiselev et~al.}(2001)\citenamefont{Kiselev, Yaroshchuk,
  Zakrevskyy, and Tereshchenko}}]{Kis:cond:2001}
\bibinfo{author}{\bibfnamefont{A.}~\bibnamefont{Kiselev}},
  \bibinfo{author}{\bibfnamefont{O.}~\bibnamefont{Yaroshchuk}},
  \bibinfo{author}{\bibfnamefont{Y.}~\bibnamefont{Zakrevskyy}},
  \bibnamefont{and}
  \bibinfo{author}{\bibfnamefont{A.}~\bibnamefont{Tereshchenko}},
  \bibinfo{journal}{Cond. Matter Phys.} \textbf{\bibinfo{volume}{4}},
  \bibinfo{pages}{67} (\bibinfo{year}{2001}).

\bibitem[{\citenamefont{Yaroshchuk
  et~al.}(2001{\natexlab{b}})\citenamefont{Yaroshchuk, Kiselev, Zakrevskyy,
  Stumpe, and Lindau}}]{Kis:epj:2001}
\bibinfo{author}{\bibfnamefont{O.}~\bibnamefont{Yaroshchuk}},
  \bibinfo{author}{\bibfnamefont{A.~D.} \bibnamefont{Kiselev}},
  \bibinfo{author}{\bibfnamefont{Y.}~\bibnamefont{Zakrevskyy}},
  \bibinfo{author}{\bibfnamefont{J.}~\bibnamefont{Stumpe}}, \bibnamefont{and}
  \bibinfo{author}{\bibfnamefont{J.}~\bibnamefont{Lindau}},
  \bibinfo{journal}{Eur. Phys. J. E} \textbf{\bibinfo{volume}{6}},
  \bibinfo{pages}{57} (\bibinfo{year}{2001}{\natexlab{b}}).

\bibitem[{\citenamefont{Hvilsted et~al.}(1992)\citenamefont{Hvilsted, Andruzzi,
  and Ramanujam}}]{Hvil:1992}
\bibinfo{author}{\bibfnamefont{S.}~\bibnamefont{Hvilsted}},
  \bibinfo{author}{\bibfnamefont{F.}~\bibnamefont{Andruzzi}}, \bibnamefont{and}
  \bibinfo{author}{\bibfnamefont{P.~S.} \bibnamefont{Ramanujam}},
  \bibinfo{journal}{Opt. Lett.} \textbf{\bibinfo{volume}{17}},
  \bibinfo{pages}{1234} (\bibinfo{year}{1992}).

\bibitem[{\citenamefont{Natansohn et~al.}(1998)\citenamefont{Natansohn, Rochon,
  Meng, Barett, Buffeteau, Bonenfant, and Pezolet}}]{Nat:1998}
\bibinfo{author}{\bibfnamefont{A.}~\bibnamefont{Natansohn}},
  \bibinfo{author}{\bibfnamefont{P.}~\bibnamefont{Rochon}},
  \bibinfo{author}{\bibfnamefont{X.}~\bibnamefont{Meng}},
  \bibinfo{author}{\bibfnamefont{C.}~\bibnamefont{Barett}},
  \bibinfo{author}{\bibfnamefont{T.}~\bibnamefont{Buffeteau}},
  \bibinfo{author}{\bibfnamefont{S.}~\bibnamefont{Bonenfant}},
  \bibnamefont{and} \bibinfo{author}{\bibfnamefont{M.}~\bibnamefont{Pezolet}},
  \bibinfo{journal}{Macromolecules} \textbf{\bibinfo{volume}{31}},
  \bibinfo{pages}{1155} (\bibinfo{year}{1998}).

\bibitem[{\citenamefont{Fisher et~al.}(1994)\citenamefont{Fisher, L\"{a}sker,
  Stumpe, and Kostromin}}]{Stu:1994}
\bibinfo{author}{\bibfnamefont{T.}~\bibnamefont{Fisher}},
  \bibinfo{author}{\bibfnamefont{L.}~\bibnamefont{L\"{a}sker}},
  \bibinfo{author}{\bibfnamefont{J.}~\bibnamefont{Stumpe}}, \bibnamefont{and}
  \bibinfo{author}{\bibfnamefont{S.~G.} \bibnamefont{Kostromin}},
  \bibinfo{journal}{J. Photochem. Photobiol. A: Chem.}
  \textbf{\bibinfo{volume}{80}}, \bibinfo{pages}{453} (\bibinfo{year}{1994}).

\bibitem[{\citenamefont{Pedersen and Michael}(1997)}]{Ped:1997}
\bibinfo{author}{\bibfnamefont{T.~G.} \bibnamefont{Pedersen}} \bibnamefont{and}
  \bibinfo{author}{\bibfnamefont{P.~M.} \bibnamefont{Michael}},
  \bibinfo{journal}{Phys. Rev. Lett.} \textbf{\bibinfo{volume}{79}},
  \bibinfo{pages}{2470} (\bibinfo{year}{1997}).

\bibitem[{\citenamefont{Pedersen et~al.}(1998)\citenamefont{Pedersen, Johansen,
  Holme, Ramanujam, and Hvilsted}}]{Ped:1998}
\bibinfo{author}{\bibfnamefont{T.~G.} \bibnamefont{Pedersen}},
  \bibinfo{author}{\bibfnamefont{P.~M.} \bibnamefont{Johansen}},
  \bibinfo{author}{\bibfnamefont{N.~C.~R.} \bibnamefont{Holme}},
  \bibinfo{author}{\bibfnamefont{P.~S.} \bibnamefont{Ramanujam}},
  \bibnamefont{and} \bibinfo{author}{\bibfnamefont{S.}~\bibnamefont{Hvilsted}},
  \bibinfo{journal}{J. Opt. Soc. Am. B} \textbf{\bibinfo{volume}{15}},
  \bibinfo{pages}{1120} (\bibinfo{year}{1998}).

\bibitem[{\citenamefont{Yaroshchuk et~al.}(1999)\citenamefont{Yaroshchuk,
  Reshetnyak, Tereshchenko, Shans'ky, Puchkovs'ka, and Lindau}}]{Puch:1999}
\bibinfo{author}{\bibfnamefont{O.~V.} \bibnamefont{Yaroshchuk}},
  \bibinfo{author}{\bibfnamefont{V.~Y.} \bibnamefont{Reshetnyak}},
  \bibinfo{author}{\bibfnamefont{A.~G.} \bibnamefont{Tereshchenko}},
  \bibinfo{author}{\bibfnamefont{L.~I.} \bibnamefont{Shans'ky}},
  \bibinfo{author}{\bibfnamefont{G.~A.} \bibnamefont{Puchkovs'ka}},
  \bibnamefont{and} \bibinfo{author}{\bibfnamefont{J.}~\bibnamefont{Lindau}},
  \bibinfo{journal}{Material Science {\upshape\&} Engineering C}
  \textbf{\bibinfo{volume}{8-9}}, \bibinfo{pages}{211} (\bibinfo{year}{1999}).

\bibitem[{\citenamefont{Sergan et~al.}(1999)\citenamefont{Sergan, Jamal, and
  Kelly}}]{Serg:1999}
\bibinfo{author}{\bibfnamefont{T.~A.} \bibnamefont{Sergan}},
  \bibinfo{author}{\bibfnamefont{S.~H.} \bibnamefont{Jamal}}, \bibnamefont{and}
  \bibinfo{author}{\bibfnamefont{J.~R.} \bibnamefont{Kelly}},
  \bibinfo{journal}{Displays} \textbf{\bibinfo{volume}{20}},
  \bibinfo{pages}{259} (\bibinfo{year}{1999}).

\bibitem[{\citenamefont{Dyadyusha et~al.}(1995)\citenamefont{Dyadyusha,
  Khizhnyak, Marusii, Reznikov, Yaroshchuk, Reshetnyak, Park, Kwon, and
  Kang}}]{Dyad:1995}
\bibinfo{author}{\bibfnamefont{A.}~\bibnamefont{Dyadyusha}},
  \bibinfo{author}{\bibfnamefont{A.}~\bibnamefont{Khizhnyak}},
  \bibinfo{author}{\bibfnamefont{T.}~\bibnamefont{Marusii}},
  \bibinfo{author}{\bibfnamefont{Y.}~\bibnamefont{Reznikov}},
  \bibinfo{author}{\bibfnamefont{O.}~\bibnamefont{Yaroshchuk}},
  \bibinfo{author}{\bibfnamefont{V.}~\bibnamefont{Reshetnyak}},
  \bibinfo{author}{\bibfnamefont{W.}~\bibnamefont{Park}},
  \bibinfo{author}{\bibfnamefont{S.}~\bibnamefont{Kwon}}, \bibnamefont{and}
  \bibinfo{author}{\bibfnamefont{D.}~\bibnamefont{Kang}},
  \bibinfo{journal}{Mol. Cryst. Liq. Cryst.} \textbf{\bibinfo{volume}{263}},
  \bibinfo{pages}{399} (\bibinfo{year}{1995}).

\bibitem[{\citenamefont{Blinov et~al.}(1983)\citenamefont{Blinov, Dubinin,
  Rumyancev, and Yudin}}]{Blin:1983}
\bibinfo{author}{\bibfnamefont{L.~M.} \bibnamefont{Blinov}},
  \bibinfo{author}{\bibfnamefont{N.~V.} \bibnamefont{Dubinin}},
  \bibinfo{author}{\bibfnamefont{V.~G.} \bibnamefont{Rumyancev}},
  \bibnamefont{and} \bibinfo{author}{\bibfnamefont{S.~G.} \bibnamefont{Yudin}},
  \bibinfo{journal}{Sov. Optika i Spectroskopiya}
  \textbf{\bibinfo{volume}{55}}, \bibinfo{pages}{679} (\bibinfo{year}{1983}),
  \bibinfo{note}{(in Russian)}.

\bibitem[{\citenamefont{Osman and Dumont}(1999)}]{Osm:1999}
\bibinfo{author}{\bibfnamefont{A.}~\bibnamefont{Osman}} \bibnamefont{and}
  \bibinfo{author}{\bibfnamefont{M.}~\bibnamefont{Dumont}},
  \bibinfo{journal}{Opt. Commun.} \textbf{\bibinfo{volume}{164}},
  \bibinfo{pages}{277} (\bibinfo{year}{1999}).

\bibitem[{\citenamefont{Feng et~al.}(1995)\citenamefont{Feng, Lin, Hooker, and
  Mickelson}}]{Feng:1995}
\bibinfo{author}{\bibfnamefont{W.}~\bibnamefont{Feng}},
  \bibinfo{author}{\bibfnamefont{S.}~\bibnamefont{Lin}},
  \bibinfo{author}{\bibfnamefont{B.}~\bibnamefont{Hooker}}, \bibnamefont{and}
  \bibinfo{author}{\bibfnamefont{A.}~\bibnamefont{Mickelson}},
  \bibinfo{journal}{Appl. Optics} \textbf{\bibinfo{volume}{34}},
  \bibinfo{pages}{6885} (\bibinfo{year}{1995}).

\bibitem[{\citenamefont{Cimrova et~al.}(1999)\citenamefont{Cimrova, Neher,
  Kostromine, and Bieringer}}]{Cim:1999}
\bibinfo{author}{\bibfnamefont{V.}~\bibnamefont{Cimrova}},
  \bibinfo{author}{\bibfnamefont{D.}~\bibnamefont{Neher}},
  \bibinfo{author}{\bibfnamefont{S.}~\bibnamefont{Kostromine}},
  \bibnamefont{and}
  \bibinfo{author}{\bibfnamefont{T.}~\bibnamefont{Bieringer}},
  \bibinfo{journal}{Macromolecules} \textbf{\bibinfo{volume}{32}},
  \bibinfo{pages}{8496} (\bibinfo{year}{1999}).

\bibitem[{\citenamefont{Kiselev}(2002)}]{Kis:jpcm:2002}
\bibinfo{author}{\bibfnamefont{A.~D.} \bibnamefont{Kiselev}},
  \bibinfo{journal}{J. Phys.: Condens. Matter} \textbf{\bibinfo{volume}{14}},
  \bibinfo{pages}{13417} (\bibinfo{year}{2002}).

\bibitem[{\citenamefont{B\"{o}hme et~al.}(1993)\citenamefont{B\"{o}hme,
  Novotna, Kresse, Kuschel, and Lindau}}]{Bohm:1993}
\bibinfo{author}{\bibfnamefont{A.}~\bibnamefont{B\"{o}hme}},
  \bibinfo{author}{\bibfnamefont{E.}~\bibnamefont{Novotna}},
  \bibinfo{author}{\bibfnamefont{H.}~\bibnamefont{Kresse}},
  \bibinfo{author}{\bibfnamefont{F.}~\bibnamefont{Kuschel}}, \bibnamefont{and}
  \bibinfo{author}{\bibfnamefont{J.}~\bibnamefont{Lindau}},
  \bibinfo{journal}{Macromol. Chem.} \textbf{\bibinfo{volume}{194}},
  \bibinfo{pages}{3341} (\bibinfo{year}{1993}).

\bibitem[{\citenamefont{Yaroshchuk
  et~al.}(2001{\natexlab{c}})\citenamefont{Yaroshchuk, Agra, Zakrevskyy, Chien,
  Lindau, and Kumar}}]{Yar:c:2001}
\bibinfo{author}{\bibfnamefont{O.}~\bibnamefont{Yaroshchuk}},
  \bibinfo{author}{\bibfnamefont{D.~M.} \bibnamefont{Agra}},
  \bibinfo{author}{\bibfnamefont{Y.}~\bibnamefont{Zakrevskyy}},
  \bibinfo{author}{\bibfnamefont{L.-C.} \bibnamefont{Chien}},
  \bibinfo{author}{\bibfnamefont{J.}~\bibnamefont{Lindau}}, \bibnamefont{and}
  \bibinfo{author}{\bibfnamefont{S.}~\bibnamefont{Kumar}},
  \bibinfo{journal}{Liq. Cryst.} \textbf{\bibinfo{volume}{28}},
  \bibinfo{pages}{703} (\bibinfo{year}{2001}{\natexlab{c}}).

\bibitem[{\citenamefont{Azzam and Bashara}(1977)}]{Azz:1977}
\bibinfo{editor}{\bibfnamefont{R.~M.~A.} \bibnamefont{Azzam}} \bibnamefont{and}
  \bibinfo{editor}{\bibfnamefont{N.~M.} \bibnamefont{Bashara}}, eds.,
  \emph{\bibinfo{title}{Ellipsometry and Polarized Light}}
  (\bibinfo{publisher}{North Holland Publishing Company},
  \bibinfo{address}{Amsterdam}, \bibinfo{year}{1977}).

\bibitem[{\citenamefont{Berreman}(1972)}]{Berr:1972}
\bibinfo{author}{\bibfnamefont{D.~W.} \bibnamefont{Berreman}},
  \bibinfo{journal}{J. Opt. Soc. Am.} \textbf{\bibinfo{volume}{62}},
  \bibinfo{pages}{502} (\bibinfo{year}{1972}).

\bibitem[{\citenamefont{Phaadt et~al.}(1996)\citenamefont{Phaadt, Boeffel, and
  Spiess}}]{Phaa:1996}
\bibinfo{author}{\bibfnamefont{M.}~\bibnamefont{Phaadt}},
  \bibinfo{author}{\bibfnamefont{C.}~\bibnamefont{Boeffel}}, \bibnamefont{and}
  \bibinfo{author}{\bibfnamefont{H.~W.} \bibnamefont{Spiess}},
  \bibinfo{journal}{Acta Polymer} \textbf{\bibinfo{volume}{47}},
  \bibinfo{pages}{35} (\bibinfo{year}{1996}).

\bibitem[{\citenamefont{Paik and Morawets}(1972)}]{Paik:1972}
\bibinfo{author}{\bibfnamefont{C.~S.} \bibnamefont{Paik}} \bibnamefont{and}
  \bibinfo{author}{\bibfnamefont{H.}~\bibnamefont{Morawets}},
  \bibinfo{journal}{Macromolecules} \textbf{\bibinfo{volume}{5}},
  \bibinfo{pages}{171} (\bibinfo{year}{1972}).

\bibitem[{\citenamefont{Eisenbach}(1980)}]{Eisen:1980}
\bibinfo{author}{\bibfnamefont{C.}~\bibnamefont{Eisenbach}},
  \bibinfo{journal}{Ber. Bunsenges. Phys. Chem.} \textbf{\bibinfo{volume}{84}},
  \bibinfo{pages}{680} (\bibinfo{year}{1980}).

\bibitem[{\citenamefont{Bernstein and Kaminskii}(1975)}]{Bern:1975}
\bibinfo{author}{\bibfnamefont{I.~Y.} \bibnamefont{Bernstein}}
  \bibnamefont{and} \bibinfo{author}{\bibfnamefont{Y.~L.}
  \bibnamefont{Kaminskii}}, \emph{\bibinfo{title}{Spectrophotometric Analysis
  in Organic Chemistry}} (\bibinfo{publisher}{Khimiya},
  \bibinfo{address}{Leningrad}, \bibinfo{year}{1975}), \bibinfo{note}{(in
  Russian)}.

\bibitem[{\citenamefont{Sajti et~al.}(2001)\citenamefont{Sajti, Kerekes,
  Barab\'{a}s, L\"{o}rincz, Hvilsted, and Ramanujam}}]{Hvil:2001}
\bibinfo{author}{\bibfnamefont{S.}~\bibnamefont{Sajti}},
  \bibinfo{author}{\bibfnamefont{A.}~\bibnamefont{Kerekes}},
  \bibinfo{author}{\bibfnamefont{M.}~\bibnamefont{Barab\'{a}s}},
  \bibinfo{author}{\bibfnamefont{E.}~\bibnamefont{L\"{o}rincz}},
  \bibinfo{author}{\bibfnamefont{S.}~\bibnamefont{Hvilsted}}, \bibnamefont{and}
  \bibinfo{author}{\bibfnamefont{P.~S.} \bibnamefont{Ramanujam}},
  \bibinfo{journal}{Opt. Commun.} \textbf{\bibinfo{volume}{194}},
  \bibinfo{pages}{435} (\bibinfo{year}{2001}).

\bibitem[{\citenamefont{Dumont}(1996)}]{Dum:1996}
\bibinfo{author}{\bibfnamefont{M.}~\bibnamefont{Dumont}}, in
  \emph{\bibinfo{booktitle}{Photoactive Organic Materials}}, edited by
  \bibinfo{editor}{\bibfnamefont{F.}~\bibnamefont{Kajzar}} \bibnamefont{et~al.}
  (\bibinfo{publisher}{Kluwer Academic Publisher},
  \bibinfo{address}{Netherlands}, \bibinfo{year}{1996}), pp.
  \bibinfo{pages}{501--511}.

\bibitem[{\citenamefont{de~Gennes and Prost}(1993)}]{Gennes:bk:1993}
\bibinfo{author}{\bibfnamefont{P.~G.} \bibnamefont{de~Gennes}}
  \bibnamefont{and} \bibinfo{author}{\bibfnamefont{J.}~\bibnamefont{Prost}},
  \emph{\bibinfo{title}{The Physics of Liquid Crystals}}
  (\bibinfo{publisher}{Claderon Press}, \bibinfo{address}{Oxford},
  \bibinfo{year}{1993}).

\bibitem[{\citenamefont{Chaikin and Lubensky}(1995)}]{Luben:bk:1995}
\bibinfo{author}{\bibfnamefont{P.~M.} \bibnamefont{Chaikin}} \bibnamefont{and}
  \bibinfo{author}{\bibfnamefont{T.~C.} \bibnamefont{Lubensky}},
  \emph{\bibinfo{title}{Principles of Condensed Matter Physics}}
  (\bibinfo{publisher}{Cambridge University Press},
  \bibinfo{address}{Cambridge}, \bibinfo{year}{1995}).

\bibitem[{\citenamefont{Abramowitz and Stegun}(1972)}]{Abr}
\bibinfo{editor}{\bibfnamefont{M.}~\bibnamefont{Abramowitz}} \bibnamefont{and}
  \bibinfo{editor}{\bibfnamefont{I.~A.} \bibnamefont{Stegun}}, eds.,
  \emph{\bibinfo{title}{Handbook of Mathematical Functions}}
  (\bibinfo{publisher}{Dover}, \bibinfo{address}{New York},
  \bibinfo{year}{1972}).

\bibitem[{\citenamefont{Mita et~al.}(1989)\citenamefont{Mita, Horie, and
  Hirao}}]{Mita:1989}
\bibinfo{author}{\bibfnamefont{I.}~\bibnamefont{Mita}},
  \bibinfo{author}{\bibfnamefont{K.}~\bibnamefont{Horie}}, \bibnamefont{and}
  \bibinfo{author}{\bibfnamefont{K.}~\bibnamefont{Hirao}},
  \bibinfo{journal}{Macromolecules} \textbf{\bibinfo{volume}{22}},
  \bibinfo{pages}{558} (\bibinfo{year}{1989}).

\bibitem[{\citenamefont{Kanazawa et~al.}(1997)\citenamefont{Kanazawa, Hirano,
  Shido, Hasegawa, Tsutsumi, Shiono, Ikeda, Nagase, Ikiyama, and
  Takamura}}]{Kanaz:1997}
\bibinfo{author}{\bibfnamefont{A.}~\bibnamefont{Kanazawa}},
  \bibinfo{author}{\bibfnamefont{S.}~\bibnamefont{Hirano}},
  \bibinfo{author}{\bibfnamefont{A.}~\bibnamefont{Shido}},
  \bibinfo{author}{\bibfnamefont{M.}~\bibnamefont{Hasegawa}},
  \bibinfo{author}{\bibfnamefont{O.}~\bibnamefont{Tsutsumi}},
  \bibinfo{author}{\bibfnamefont{T.}~\bibnamefont{Shiono}},
  \bibinfo{author}{\bibfnamefont{T.}~\bibnamefont{Ikeda}},
  \bibinfo{author}{\bibfnamefont{Y.}~\bibnamefont{Nagase}},
  \bibinfo{author}{\bibfnamefont{E.}~\bibnamefont{Ikiyama}}, \bibnamefont{and}
  \bibinfo{author}{\bibfnamefont{Y.}~\bibnamefont{Takamura}},
  \bibinfo{journal}{Liq. Cryst.} \textbf{\bibinfo{volume}{23}},
  \bibinfo{pages}{293} (\bibinfo{year}{1997}).

\end{thebibliography}

\end{document}